\documentclass[%
 reprint,
superscriptaddress,
 amsmath,amssymb,
 aps,
 prl,
]{revtex4-2}
\usepackage{placeins}
\usepackage{graphicx}
\usepackage{dcolumn}
\usepackage{bm}
\usepackage{physics} 
\usepackage{hyperref}
\usepackage{mathtools}

\usepackage[caption=false]{subfig}

\begin{document}

\preprint{APS/123-QED}

\title{Multimode Phonon-Number Measurement and Single-shot Superparity Measurement using Dispersive Shifts in a Trapped Ion}

\author{Wonhyeong Choi}
\thanks{These two authors contributed equally}
\affiliation{Department of Computer Science and Engineering, Seoul National University}
\affiliation{Automation and Systems Research Institute, Seoul National University}
\affiliation{NextQuantum, Seoul National University}

\author{Jiyong Kang}
\thanks{These two authors contributed equally}
\affiliation{Department of Computer Science and Engineering, Seoul National University}
\affiliation{Automation and Systems Research Institute, Seoul National University}
\affiliation{NextQuantum, Seoul National University}

\author{Kyunghye Kim}
\affiliation{Department of Computer Science and Engineering, Seoul National University}
\affiliation{Automation and Systems Research Institute, Seoul National University}
\affiliation{NextQuantum, Seoul National University}

\author{Jaehun You}
\affiliation{Department of Computer Science and Engineering, Seoul National University}
\affiliation{Automation and Systems Research Institute, Seoul National University}
\affiliation{NextQuantum, Seoul National University}

\author{Kyungmin Lee}
\affiliation{Department of Computer Science and Engineering, Seoul National University}
\affiliation{Automation and Systems Research Institute, Seoul National University}
\affiliation{NextQuantum, Seoul National University}

\author{Taehyun Kim}
\email{taehyun@snu.ac.kr}
\affiliation{Department of Computer Science and Engineering, Seoul National University}
\affiliation{Automation and Systems Research Institute, Seoul National University}
\affiliation{NextQuantum, Seoul National University}
\affiliation{Institute of Applied Physics, Seoul National University}
\date{\today}

\begin{abstract}
Dispersive shifts are a widely used tool for bosonic readout and control in circuit quantum electrodynamics, yet they remain relatively unexplored in trapped-ion motional systems. Here we introduce a unified framework for multimode phonon-number measurement and nondestructive single-shot superparity measurement, i.e., phonon-number measurement modulo $2^k$, using dispersive shifts in the far-detuned multimode Jaynes-Cummings interaction of a trapped ion system. We implement a Ramsey sequence that realizes a multimode spin-dependent rotation ($\operatorname{SDR}$) together with a selective decoupling scheme that cancels the phase induced by the carrier AC-Stark shift while preserving the phonon-number-dependent phase induced by the dispersive shift. Within this framework, we infer single-mode and two-mode Fock-state distributions from spin-population dynamics, use $\operatorname{SDR}$-based conditional parity operators with postselection to generate cat states and entangled coherent states, and realize nondestructive single-shot measurements of phonon number modulo $2$, $4$, and $8$ in the single-mode setting. These results open a new avenue for the use of multimode parity operators in trapped-ion bosonic systems.
\end{abstract}

\maketitle
\section{Introduction}
Bosonic modes provide powerful alternatives to qubits in quantum information processing~\cite{braunsteinQuantumInformationContinuous2005b}. While a single qubit is confined to a two-dimensional Hilbert space, a single bosonic mode provides an infinite-dimensional Hilbert space, enabling continuous-variable quantum computation~\cite{lloydQuantumComputationContinuous1999, michaelNewClassQuantum2016, gottesmanEncodingQubitOscillator2001}, hybrid quantum computation~\cite{andersenHybridDiscreteContinuousvariable2015, liuHybridOscillatorQubitQuantum2024}, quantum metrology~\cite{giovannettiAdvancesQuantumMetrology2011,mccormickQuantumenhancedSensingSingleion2019,dengQuantumenhancedMetrologyLarge2024,millicanEngineeringContinuousvariableEntanglement2025,valahuQuantumenhancedMultiparameterSensing2025, bondOptimalDisplacementSensing2025}, and quantum simulation~\cite{georgescuQuantumSimulation2014, lvQuantumSimulationQuantum2018a, dengObservingQuantumTopology2022, craneHybridOscillatorQubitQuantum2024}. These ideas are being pursued across circuit quantum electrodynamics (cQED)~\cite{puttermanHardwareefficientQuantumError2025, dengObservingQuantumTopology2022, dengQuantumenhancedMetrologyLarge2024}, neutral atoms~\cite{leongCreationTwoModeSqueezed2023,brownTimeofflightQuantumTomography2023, shawErasureCoolingControl2025}, single photons~\cite{larsenIntegratedPhotonicSource2025, jeongGenerationHybridEntanglement2014a, leeFaultTolerantQuantumComputation2024,leePhotonicHybridQuantum2025}, and trapped ions~\cite{mccormickQuantumenhancedSensingSingleion2019,fluhmannEncodingQubitTrappedion2019, matsosRobustDeterministicPreparation2024, matsosUniversalQuantumGate2025,ganHybridQuantumComputing2020}. Trapped ions offer a natural setting for multimode bosonic registers: long-lived internal states serve as qubits while collective vibrational modes provide bosonic resources. Specifically, an $N$-ion crystal supports $3N$ motional modes, so a large bosonic register naturally coexists and scales with the qubit register, enabling multimode bosonic encodings~\cite{nohEncodingOscillatorMany2020,royerEncodingQubitsMultimode2022}.

The characterization of such bosonic states is a key ingredient in bosonic quantum information processing.
In trapped-ion platforms, the single-mode case has been studied extensively, ranging from phase-space tomography via characteristic functions~\cite{fluhmannDirectCharacteristicFunctionTomography2020} or Wigner functions~\cite{ganHybridQuantumComputing2020} to Fock-state population estimation using sideband spectroscopy~\cite{leibfriedExperimentalDeterminationMotional1996} or cross-Kerr nonlinearities~\cite{dingQuantumParametricOscillator2017b,dingCrossKerrNonlinearityPhonon2017a}, as well as nondestructive single-shot measurements of individual Fock states~\cite{mallwegerSingleShotMeasurementsPhonon2023}. 
A few multimode demonstrations exist, including two-mode characteristic functions~\cite{whitlowQuantumSimulationConical2023,valahuDirectObservationGeometricphase2023, millicanEngineeringContinuousvariableEntanglement2025}, two-mode Wigner functions~\cite{jeonMultimodeBosonicState2025}, and population estimation using multimode sideband spectroscopy~\cite{jiaDeterminationMultimodeMotional2022}. While these trapped-ion measurement schemes provide a rich toolbox, they do not cover all situations of practical interest. In particular, multimode sideband spectroscopy becomes challenging for population measurement when the number of motional modes exceeds the number of available ions~\cite{jiaDeterminationMultimodeMotional2022}. At the same time, measurements of phonon number modulo $N$, relevant to bosonic quantum error correction~\cite{michaelNewClassQuantum2016,ofekExtendingLifetimeQuantum2016,grimsmoQuantumComputingRotationSymmetric2020}, have not yet been investigated for trapped-ion motional modes.

In the dispersive regime of the Jaynes-Cummings Hamiltonian, the interaction reduces to a phonon- or photon-number-dependent frequency shift, or equivalently to a spin-dependent rotation ($\operatorname{SDR}$) in phase space~\cite{larsonJaynesCummingsModelIts2021}. Dispersive shifts have played a central role in cQED, enabling high-fidelity readout~\cite{blaisCavityQuantumElectrodynamics2004, schusterAcStarkShift2005,wallraffApproachingUnitVisibility2005a,wallraffStrongCouplingSingle2004}, nonclassical bosonic state generation~\cite{vlastakisDeterministicallyEncodingQuantum2013a,wangConvertingQuasiclassicalStates2017,dengQuantumenhancedMetrologyLarge2024, sunTrackingPhotonJumps2014}, and universal bosonic control~\cite{heeresImplementingUniversalGate2017,krastanovUniversalControlOscillator2015,schusterResolvingPhotonNumber2007,eickbuschFastUniversalControl2022}, and have also been exploited in quantum acoustics~\cite{wollackQuantumStatePreparation2022}. Dispersive shifts connect naturally to parity-based observables such as the Wigner function which is the expectation value of a displaced parity operator~\cite{royerWignerFunctionExpectation1977}. In the two-mode extension of the parity operator, the corresponding joint-parity gives access to joint Wigner functions~\cite{wangSchrodingerCatLiving2016,jeonMultimodeBosonicState2025} and joint-$4$ parity operators can serve as entangling operators between bosonic qubits~\cite{tsunodaErrorDetectableBosonicEntangling2023}. 
Two-mode number difference observables can also define useful nonclassical manifolds such as pair-coherent states and pair-cat states~\cite{gertlerExperimentalRealizationCharacterization2023,albertPaircatCodesAutonomous2019}. So far, trapped-ion experiments have explored dispersive shifts as single-mode spin-motion entanglement~\cite{schmidt-kalerQuantizedACStarkShifts2004}, cavity-photon population probing~\cite{leeIonBasedQuantumSensor2019}, and conditional number gates~\cite{valahuQuantumenhancedMultiparameterSensing2025}. However, experimental studies that use dispersive shifts as a tool to generalize parity operators in trapped ions remain unexplored.

Here, we introduce a unified framework for multimode phonon-number measurement and nondestructive single-shot superparity measurement, i.e., phonon-number measurement modulo $2^k$~\cite{liuHybridOscillatorQubitQuantum2024}, in a trapped-ion system using dispersive shifts in the multimode Jaynes-Cummings interaction. We implement a Ramsey sequence that realizes a multimode $\operatorname{SDR}$ together with a selective decoupling scheme that cancels the phase induced by the carrier AC-Stark shift while preserving the phonon-number-dependent phase induced by the dispersive shift. 
This $\operatorname{SDR}$-based framework enables two complementary measurement capabilities. First, by fitting the spin-population dynamics, we infer single-mode and two-mode Fock-state distributions. We also use $\operatorname{SDR}$-based parity filtering of coherent states to generate cat states and entangled coherent states (ECSs), whose phonon-number distributions are inferred within the same framework. Second, by sequentially applying the $\operatorname{SDR}$ protocol, we realize nondestructive single-shot phonon-number measurements modulo $2$, $4$, and $8$ in the single-mode setting.

\section{Results}
\subsection{Multimode dispersive shift and spin-dependent rotation}
We consider a single spin coupled to $M$ motional modes and driven by a single off-resonant laser tone on either the red or blue motional sidebands. The laser frequency is detuned from the red (blue) sideband of mode $j$ by $\delta_j$ and hence from the carrier transition by $\Delta = \mp\omega_j + \delta_j$, where $\omega_j$ is the secular frequency of mode $j$ and the upper (lower) sign corresponds to the red (blue) sideband. Within the rotating-wave approximation, this single-tone laser couples the spin to all motional modes, and the interaction Hamiltonian takes the multimode Jaynes-Cummings form
\begin{equation}
\hat{H}_I(t)=\hbar\sum_{j=1}^M \frac{\eta_j \Omega}{2}\Big( \hat{\sigma}_{\pm}\hat{a}_j e^{\mp i\delta_j t}+\hat{\sigma}_{\mp}\hat{a}_j^\dagger e^{\pm i\delta_j t}\Big),
\label{eq:multimodeJC}
\end{equation}
where $\eta_j$ is the Lamb-Dicke parameter for mode \(j\), and \(\Omega\) is the Rabi frequency of the carrier transition. In the limit of \(|\delta_j|\!\gg\! \eta_j |\Omega|\sqrt{n_j+1}\) and using the fact that $|\delta_i-\delta_j|=|\omega_i-\omega_j|$ is sufficiently large for any pair of $(i, j)$, the Schrieffer-Wolff transformation yields the effective Hamiltonian~\cite{SW},
\begin{equation}
\hat{H}_{\mathrm{disp}}
\simeq\hbar\,\hat{\sigma}_z\!\sum_{j=1}^M\chi_j\!\left(\hat{n}_j+\tfrac12\right),
\label{eq:H-dispersive-shift}
\end{equation}
with coefficients
\begin{equation}
\chi_j=-\frac{\eta_j^2 \Omega^2}{4\,\delta_j},
\label{eq:params_twomode_hat_1}
\end{equation}
where \(\hat{n}_j= \hat{a}_j^\dagger \hat{a}_j\). Note that when the sidebands are crowded or when a multi-tone drive is applied, a Schrieffer-Wolff transformation yields a spin-dependent beam-splitter term in the effective Hamiltonian~\cite{chenScalableProgrammablePhononic2023,katzProgrammableQuantumSimulations2023}. In our system, the motional modes are well separated, so the beam-splitter term itself is sufficiently off-resonant to be neglected in the following (See Supplementary Information).

We use the dispersive-shift interaction in Eq.~(\ref{eq:H-dispersive-shift}) to encode phonon-number-dependent phases on the spin state. The resulting unitary operator is a multimode $\operatorname{SDR}$,
\begin{equation}
\operatorname{SDR}(\boldsymbol{\theta})
\coloneqq e^{-i \hat{\sigma}_z\,\boldsymbol{\theta}\cdot \boldsymbol{\hat n}/2},
\label{eq:gen_parity}
\end{equation}
where $\boldsymbol{\theta}=(\theta_1,\theta_2,\ldots,\theta_M)$ and $\boldsymbol{\hat n}=(\hat n_1,\hat n_2,\ldots,\hat{n}_M)$, up to a residual $\hat \sigma_z$ rotation $e^{-i \hat{\sigma}_z\,\phi_\mathrm{off}/2}$ with $\phi_\mathrm{off} = \sum_j \theta_j/2$. The parameter vector $\boldsymbol{\theta}$ is controlled by the laser detunings and interaction times. For $M=2$, choosing $\boldsymbol{\theta}=(\pi,0)$ implements the spin-dependent single-mode parity operator used in Wigner-function measurements and parity-based metrology~\cite{royerWignerFunctionExpectation1977,birrittellaParityOperatorApplications2021,gerryParityOperatorQuantum2010}. More generally, $\boldsymbol{\theta}=(\pi,\pi)$ and $(\pi/2,\pi/2)$ realize the spin-dependent joint-parity and joint-4-parity operators, respectively, which underlie two-mode Wigner tomography~\cite{wangSchrodingerCatLiving2016,jeonMultimodeBosonicState2025} and controlled-$ZZ$ gates between bosonic qubits such as binomial- or cat-code states~\cite{tsunodaErrorDetectableBosonicEntangling2023}.

To access the multimode Fock-state distribution, we embed $\operatorname{SDR}$ in a Ramsey sequence. 
We first prepare $\hat\rho=\ket \downarrow\bra \downarrow\otimes\hat\rho_{\mathrm{mot}}$ where $\hat\rho_{\mathrm{mot}}$ is the density operator of the $M$-mode motional system.
Define the multimode Fock-state populations
\begin{equation}
p_{\boldsymbol n}\coloneqq \bra{\boldsymbol n}\hat\rho_{\mathrm{mot}}\ket{\boldsymbol n},
\label{eq:pn_def_app}
\end{equation}
where $\ket{\boldsymbol{n}} = \ket{n_1, n_2,\ldots, n_M}$. 
We apply a $\pi/2$ pulse to prepare the spin state in a superposition of $\hat\sigma_z$ eigenstates, after which the $\operatorname{SDR}$ imprints opposite phonon-number-dependent phases on the two spin components. A second $\pi/2$ rotation is then applied, and we measure the $\ket{\uparrow}$-state population. The full sequence yields
\begin{equation}
P_{\uparrow}(t) 
= \tfrac{1}{2} -\tfrac{1}{2}\sum_{\boldsymbol{n}\in\mathbb{N}_0^M} p_{\boldsymbol{n}}
   \cos\!\left(\boldsymbol{\theta}(t)\cdot \boldsymbol{n} \right),
\label{eq:Pr-up-cos}
\end{equation}
where $\boldsymbol{n}=(n_1,n_2,\ldots,n_M)$. By fitting Eq.~(\ref{eq:Pr-up-cos}) to the measured $P_\uparrow(t)$, we determine the multimode Fock-state distribution $p_{\boldsymbol{n}}$.

\subsection{Experimental system}
In the experiments, we use two radial motional modes and a two-level spin state of a single $^{171}\mathrm{Yb}^{+}$ ion confined in a linear Paul trap with blade-shaped electrodes~\cite{jeonExperimentalRealizationEntangled2024,jeonMultimodeBosonicState2025}. The secular frequencies of the two radial modes are $\omega_1 / (2\pi) \approx 0.94 \ \mathrm{MHz}$ and $\omega_2 / (2\pi) \approx 1.27 \ \mathrm{MHz}$. The motional modes are coupled to the spin states $\ket{\downarrow} \coloneqq \ket{F=0,\,m_F=0}$ and $\ket{\uparrow} \coloneqq \ket{F=1,\,m_F=0}$ of the $S_{1/2}$ manifold by stimulated Raman transitions with pulsed 355-nm beams, yielding Lamb-Dicke parameters $\eta_1 = 0.10$ and $\eta_2 = 0.087$, while pure spin rotations are implemented with a resonant microwave field to realize high-fidelity $\pi$ and $\pi/2$ pulses. State preparation and detection are performed via optical pumping to $\ket{\downarrow}$ and state-dependent fluorescence at 369.5 nm, respectively, for which $\ket{\downarrow}$ is dark~\cite{olmschenkManipulation2007}. To implement the dispersive shift, we off-resonantly drive the first-order red- or blue-sideband transition of the radial motional modes.

\subsection{Direct spectroscopy of the dispersive shift}
\begin{figure*}[t]
\includegraphics[width=0.9\textwidth]{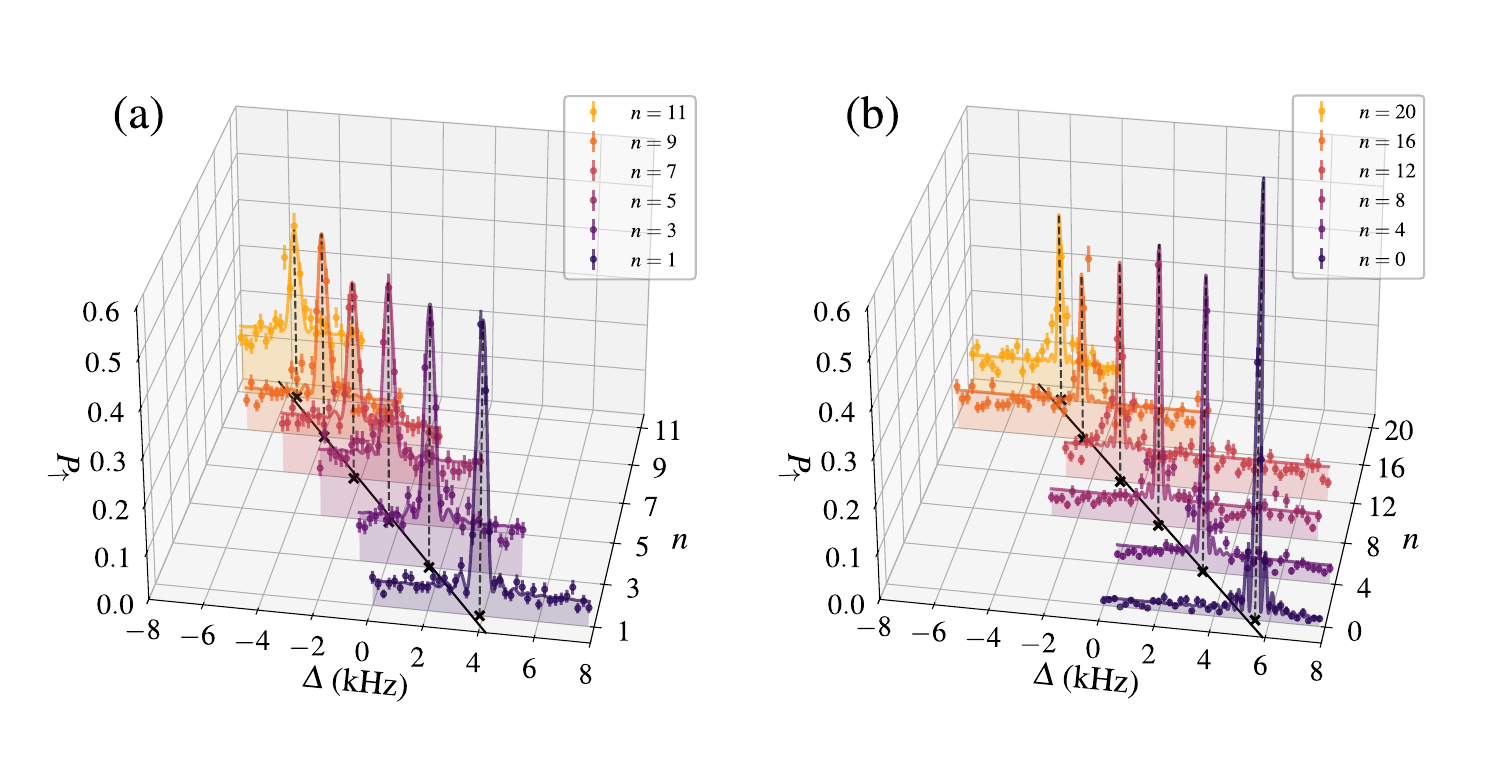}
\caption{Direct spectroscopy of dispersive shifts using microwave in a single trapped-ion qubit and motional state prepared in individual Fock state $\ket{n}$ of mode~1. A Raman coupling field is applied detuned from the red sideband of mode~1 to generate a dispersive shift, while a weak microwave probe scans the qubit transition. We scan the microwave detuning 
$\Delta$ relative to the qubit transition frequency measured in the absence of the Raman beam. The probe amplitude is chosen to yield a linewidth narrow enough to resolve adjacent $n$. The Raman amplitude is calibrated to give a carrier Rabi rate of $100~\mathrm{kHz}$. For each detuning scan result, we fit the data to find the center of the peak. The fitted centers are denoted by black cross markers on the $P_\uparrow=0$ plane. 
We then extract the phonon-number-dependent shift from a linear fit, shown as a black line on the same plane. (a) $50~\mathrm{kHz}$ from the red sideband and for a duration of $2~\mathrm{ms}$. From the linear fit, we get $\chi_{\mathrm{eff}, 1} / (2\pi)=-471(15)~\mathrm{Hz}$. (b) $100~\mathrm{kHz}$ from the red sideband and for duration of $4~\mathrm{ms}$. From the linear fit, we get $\chi_{\mathrm{eff}, 1} / (2\pi)=-246(9)~\mathrm{Hz}$.}
\label{fig:direct_measure}
\end{figure*}

By driving the motional red sideband off-resonantly, the qubit experiences a dispersive shift in the regime $\eta\Omega \ll |\delta|$ as described by Eq.~\eqref{eq:H-dispersive-shift}. This produces an approximately linear energy shift proportional to the phonon number $n$, which we read out by scanning the microwave frequency near the qubit resonance. We prepare a single-mode Fock state, apply a Raman field as a coupling field to generate the dispersive shift, and use a weak microwave field (probe field) to scan the qubit transition with a linewidth narrow enough to resolve individual $n$. For the data in Fig.~\ref{fig:direct_measure}, the Raman amplitude was calibrated to yield a carrier Rabi rate of $\Omega/(2\pi)= 100~\mathrm{kHz}$. We tested two settings: detunings of $50~\mathrm{kHz}$ and $100~\mathrm{kHz}$ from the red sideband of mode 1 with microwave probe times of $2~\mathrm{ms}$ and $4~\mathrm{ms}$, respectively. Both conditions produce dispersive shifts that increase with $n$ as shown in Fig.~\ref{fig:direct_measure}. At larger Fock numbers, where $\eta\Omega\sqrt{n}$ becomes comparable to $|\delta|$, residual sideband excitation produces a background offset in the scans, which grows with the Fock number as shown in Fig.~\ref{fig:direct_measure}. This approach is similar to the single-shot Fock-state measurement in Refs.~\cite{mallwegerSingleShotMeasurementsPhonon2023, dingCrossKerrNonlinearityPhonon2017a}. Using this scheme, direct frequency scans may provide an alternative route to reconstruct the Fock-state distribution and could enable single-shot Fock-number readout and selective number-dependent arbitrary phase (SNAP) gates~\cite{heeresImplementingUniversalGate2017, krastanovUniversalControlOscillator2015,eickbuschFastUniversalControl2022}. A limitation is sensitivity to laser-intensity noise due to AC-Stark shift, but this may be mitigated by adding a compensating tone to cancel the AC-Stark shift induced by the carrier~\cite{haffnerPrecisionMeasurementCompensation2003}. In the following, instead, we compensate the noise from AC-Stark shift using spin-echo type cancellation.

\subsection{Selective decoupling for $\operatorname{SDR}$}
A single-tone laser detuned from the motional sideband produces a dispersive shift but also drives the carrier transition off-resonantly, inducing an AC-Stark shift,
$\hat{H}_{\mathrm{Stark}}=-\frac{\hbar\,\Omega^{2}}{4\Delta}\,\hat{\sigma}_{z},$
where $\Omega$ is the carrier Rabi frequency and $\Delta$ is the detuning from the carrier. The total effective Hamiltonian under this drive is therefore
\begin{equation}
\hat{H}_{\mathrm{tot}} = \hat{H}_{\mathrm{disp}} + \hat{H}_{\mathrm{Stark}}.
\end{equation}
Using $\eta^{2} \approx 10^{-2}$ and $\delta/\Delta \approx 10^{-1}$, we estimate that the dispersive shift $\chi \propto \eta^{2}\Omega^{2}/\delta$ is about one order of magnitude smaller than the carrier AC-Stark shift. As a result, fluctuations of the laser intensity generate spin-phase noise comparable to the desired phonon-number-dependent phase, making this phase difficult to resolve experimentally. Previous work has canceled the carrier AC-Stark shift using an additional compensation tone~\cite{haffnerPrecisionMeasurementCompensation2003, schmidt-kalerQuantizedACStarkShifts2004}. Here, we instead introduce a selective decoupling scheme that cancels the phase induced by the carrier AC-Stark shift while preserving the phase induced by the dispersive shift.

\begin{figure}[t]
\includegraphics[width=\columnwidth]{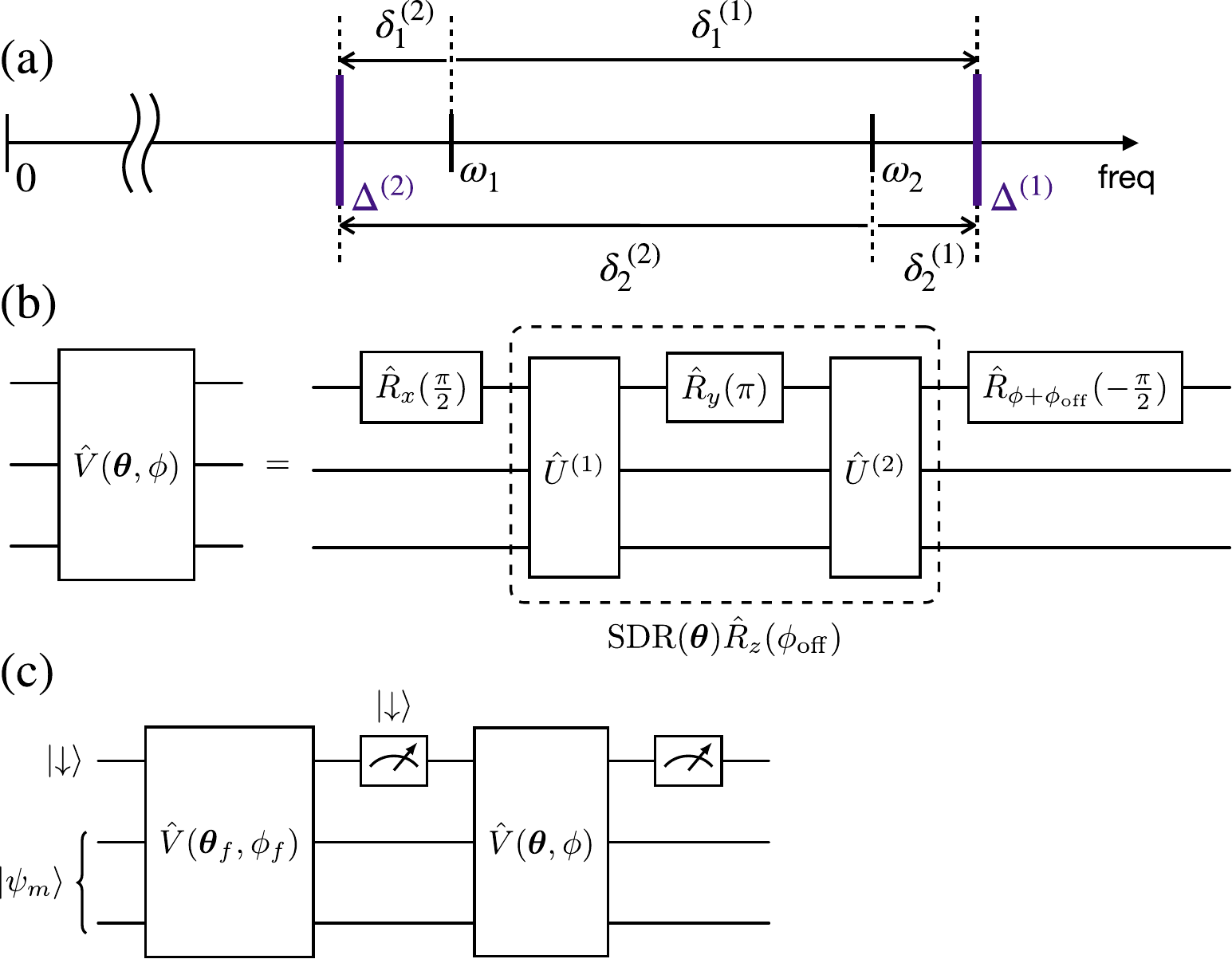}
\caption{\label{fig1}
(a) Laser detunings and sideband frequencies relative to the carrier transition in the two-mode $\operatorname{SDR}$. The sign of detuning relative to both sidebands is flipped in step $2$ to make the phonon-number-dependent phase constructive.
(b) Ramsey sequence with selective decoupling used to apply dispersive shifts for filtering and measurement. The sequence is divided into two step segments separated by a microwave $\pi$ pulse, which acts as a spin echo. This cancels the carrier-induced AC-Stark shift while preserving the dispersive shift. 
(c) Sequence of the parity-based filtering and Fock-state population measurement. The protocol in (b) is applied twice, where the first run with a measurement implements the filtering and the second extracts the Fock-state populations.
}
\end{figure}

A single-tone laser drive generates the unitary operator
\begin{equation}
\hat{U}(\Delta,t)\coloneqq \exp\!\Big[-\tfrac{i}{\hbar}\,\hat{H}_{\mathrm{tot}}(\Delta)\,t\Big].
\end{equation}
To implement a two-mode $\operatorname{SDR}(\boldsymbol{\theta})$ with selective decoupling, we insert a microwave $\pi$ pulse as a spin echo between two consecutive off-resonant single-tone interactions. Denoting the evolution in step $k$ by $\hat{U}^{(k)} \coloneqq \hat{U}(\Delta^{\!(k)},t^{(k)})$ where the superscript denotes the step index and $\hat R_\alpha(\theta) \coloneqq e^{-i\hat\sigma_\alpha\theta/2}$ for $\alpha\in\{x,y,z,\phi\}$ with $\hat\sigma_\phi \coloneqq \cos\phi\,\hat\sigma_x+\sin\phi\,\hat\sigma_y$, the total unitary operator of the two-step sequence is
\begin{equation}
\hat{U}^{(2)}\hat{R}_y(\pi)\,\hat{U}^{(1)}
=\operatorname{SDR}(\boldsymbol{\theta})\,\hat R_z(\phi_{\mathrm{off}}).
\label{eq:parity_target}
\end{equation}
The off-resonant carrier transition in each step segment induces an AC-Stark shift, imparting a phase on the spin,
\begin{equation}
\phi_{\mathrm{Stark}} \propto \frac{\Omega^2}{4}\Big(\frac{t^{(1)}}{\Delta^{\!(1)}}-\frac{t^{(2)}}{\Delta^{\!(2)}}\Big),
\end{equation}
where the relative minus sign arises from the $\hat R_y(\pi)$ flip between the two segments. Selective decoupling from this AC-Stark shift is achieved by imposing ${t^{(1)}}/{t^{(2)}}={\Delta^{\!(1)}}/{\Delta^{\!(2)}}$
so that $\phi_{\mathrm{Stark}}$ vanishes. 
Choosing detunings on opposite sides of the relevant sideband(s) across the intermediate spin flip makes the phonon-number-dependent phases add constructively. The accumulated phase for mode $j$ is $\theta_j=2(-\chi_j^{(1)}t^{(1)}+\chi_j^{(2)}t^{(2)})$. We define the effective dispersive shift as $\chi_{\mathrm{eff},j}\coloneqq\sum_k(-1)^k\chi_j^{(k)}t^{(k)}/\sum_{k'} t^{(k')}$, so that $\theta_j=2\chi_{\mathrm{eff},j}\sum_k t^{(k)}$ by definition. 
For two modes, we choose the first detuning above the higher sideband and the second below the lower sideband as illustrated in Fig.~\ref{fig1}(a), so that both modes acquire constructive phonon-number-dependent phases. For a single mode, we place the two detunings on opposite sides of the sideband of the selected mode. 
The selective-decoupling experiments reported below use the blue sideband, although the protocol itself works with either the red or blue sideband.

We combine two $\pi/2$ pulses to form a Ramsey sequence with selective decoupling and denote the resulting operation by
\begin{equation} \label{eq:V-operator}
    \hat{V}(\boldsymbol{\theta},\phi)=\hat R_{\phi+\phi_{\mathrm{off}}}(-\frac \pi 2)\hat{U}^{(2)}\hat{R}_y(\pi)\,\hat{U}^{(1)}\hat R_x(\frac \pi 2),
\end{equation}
as illustrated in Fig.~\ref{fig1}(b). Because the dispersive shift scales as $\chi_{j} (n_j + \tfrac{1}{2})$, the accumulated phase carries a phonon-number-independent offset. Our Ramsey sequence with selective decoupling isolates the desired phonon-number-dependent phase, while a residual $\hat \sigma_z$ rotation from the $+\tfrac{1}{2}$ term remains. We cancel this trivial offset by advancing the phase of the second microwave $\pi/2$ pulse by $\phi_{\mathrm{off}} = \sum_j \theta_j/2$ which implements an effective $\hat \sigma_z$ rotation $\hat R_z(-\phi_{\mathrm{off}})$, so that the net phase is strictly proportional to $\sum_j \chi_{\mathrm{eff},j} n_j$ (See Methods).

\subsection{Linearity of the dispersive shift}
\begin{figure}
    \centering
    \includegraphics[width=\columnwidth]{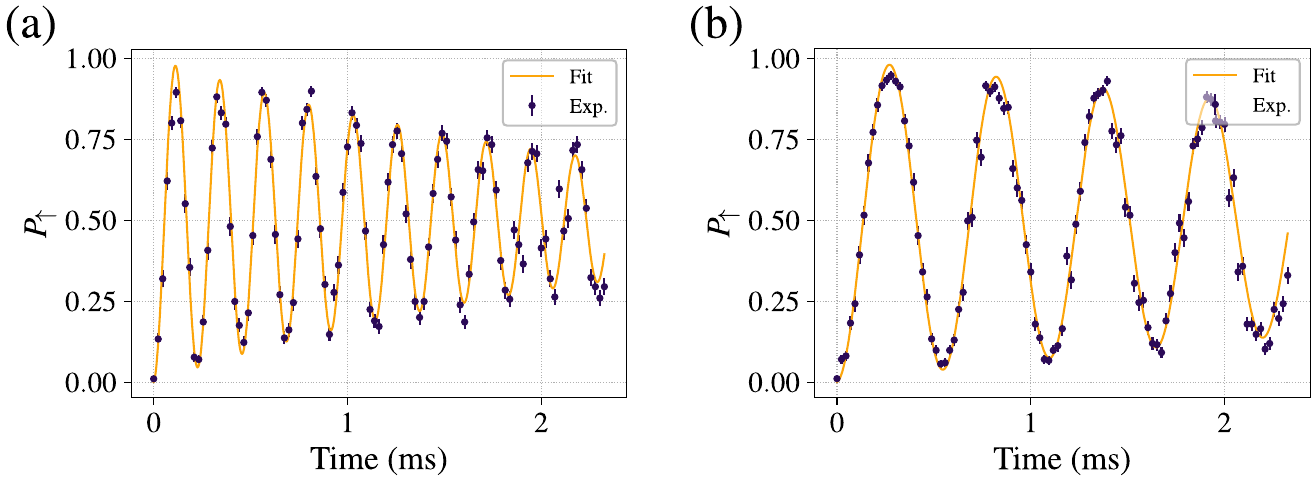}
    \caption{Oscillation of the spin population with an exponential decay of the contrast for the single-mode Fock states (a) $\ket{10}$ and (b) $\ket{4}$. The markers show the experimental data and the curves show the fitted $f_\uparrow (t)$ defined in Eq.~\eqref{eq:f_uparrow}. The error bars are calculated from the quantum projection noise over 300 repetitions.}
    \label{fig:fock-time-evolution}
\end{figure}

We first characterized the $\operatorname{SDR}$ Ramsey sequence using single-mode and multimode Fock states. 
As Eq.~\eqref{eq:Pr-up-cos} implies, the spin population exhibits sinusoidal oscillations for a pure Fock state. However, in the measured data we observe a decay of the oscillation amplitude. We model the decay of the contrast as exponential and extract the decay constant $\gamma$ by fitting the oscillations to the following function:
\begin{equation} \label{eq:f_uparrow}
    f_\uparrow (t; \gamma, \chi, \phi) = \frac 1 2 \left( 1 - e^{-\gamma t} \cos (2\chi t + \phi) \right).
\end{equation}
In Fig.~\ref{fig:fock-time-evolution} we show the time evolutions of the spin population $P_\uparrow (t)$ for the single-mode Fock states $\ket{10}$ and $\ket{4}$. The experimental data markers show good agreement with the model curve, showing that the $f_\uparrow$ function effectively captures the decay.
\begin{figure*}
    \centering
    \includegraphics[width=\textwidth]{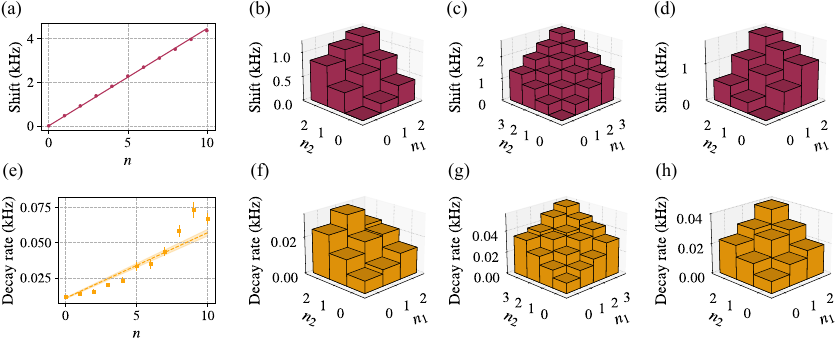}
    \caption{
    (a)--(d) Correlation between the effective dispersive shift and the phonon number, and (e)--(h) correlation between the decay rate and the phonon number. 
    Panels (a) and (e) correspond to the single-mode setting for mode~1, for which a linear fit gives $\chi_{\mathrm{eff},1}/(2\pi)=-222.8(2)\,\mathrm{Hz}$. 
    Panels (b, f), (c, g), and (d, h) correspond to multimode settings with target effective dispersive-shift ratios $\chi_{\mathrm{eff},1}/\chi_{\mathrm{eff},2}=0.5$, $1$, and $2$, respectively. From linear fits to the data in (b)--(d), we obtain $(\chi_{\mathrm{eff},1},\chi_{\mathrm{eff},2})/(2\pi)=(-103.8(5),-207.6(6))\,\mathrm{Hz}$ for the $1\!:\!2$ setting, $(-233.4(5),-224.8(5))\,\mathrm{Hz}$ for the $1\!:\!1$ setting, and $(-270.8(6),-135.1(5))\,\mathrm{Hz}$ for the $2\!:\!1$ setting. The bars (markers in (a,e)) show the fitted $f_\uparrow$-function parameters---$2\chi$ for the shift and $\gamma$ for the decay rate---extracted from the experimental data. The solid line in (a) and the dashed line in (e) show the linear fits, where the shaded areas indicate the fitting errors. The error bars in (a, e) show the quantum projection noise over 300 repetitions. The shaded area and the error bars in (a) are not visible because of the small uncertainties.}
    \label{fig:linearity}
\end{figure*}
In Fig.~\ref{fig:linearity} we measure the effective dispersive shift $2\chi$ and the decay rate $\gamma$ for a variety of Fock states, except for the motional vacuum state $\ket{0}$, for which we set $\chi=0$ without a sinusoidal fit, because we suppress the spin oscillation for the vacuum state by calibration. The total dispersive shift then can be described by a linear combination of the phonon numbers, $2\chi = 2\chi_{\mathrm{eff, 1}} n_1 + 2\chi_{\mathrm{eff, 2}}n_2$, where $n_j$ is the phonon number in mode~$j$. The fitted values of the effective dispersive shift shown in Figs.~\ref{fig:linearity}(a)--(d) are described in the Supplementary Information. In Figs.~\ref{fig:linearity}(e)--(h) the decay coefficient $\gamma$ is similarly correlated with the phonon numbers, in agreement with previous work in quantum acoustodynamics~\cite{wollackQuantumStatePreparation2022}. However, $\gamma$ drifts over time in practice, so the phonon-number dependence shows deviations from the linear trends. Because of this drift, we fit the decay rate individually for every experimental dataset in this work.

\subsection{Determination of multimode phonon-number distribution of parity-filtered states}
Applying the Ramsey sequence gives access to the Fock-state populations through Eq.~(\ref{eq:Pr-up-cos}). In the single-mode case, $P_\uparrow(t)$ is strictly periodic. In the multimode case, however, the dynamics are generally not periodic when several dispersive shifts $\{\chi_{\mathrm{eff},j}\}$ contribute with irrational ratios. To determine the multimode Fock-state populations, one can proceed in two ways: (\romannumeral 1) choose several settings where the ratios $\chi_{\mathrm{eff},i}/\chi_{\mathrm{eff},j}$ are rational and observe time evolutions of $P_\uparrow(t)$ for each setting, so that every data set is periodic and can be fitted simultaneously, or (\romannumeral 2) choose a setting where the ratios are irrational and observe a single long time evolution of $P_\uparrow(t)$ that contains sufficient frequency information to resolve all components~\cite{wollackQuantumStatePreparation2022}. Given the motional coherence time of a few ms in our setup, we adopt strategy~(\romannumeral 1).

Using the sequence of Fig.~\ref{fig1}(c), we implement parity-based filtering by choosing appropriate $\boldsymbol{\theta}_f$ to project onto the desired parity subspace. After the spin state is initialized to $\ket{\downarrow}$, we apply $\hat V(\boldsymbol{\theta}_f,\phi_f)$ and then measure the spin to collapse the spin-motion entangled state. By postselecting on $\ket{\downarrow}$, the dark state, we obtain the motional state with the desired parity. We choose $\boldsymbol{\theta}_f=(\pi,0)$ and $(\pi,\pi)$ to realize the single-mode parity and the joint-parity operators, respectively. Whether $\phi_f = 0$ or $\phi_f = \pi$ determines which parity sector---even or odd, respectively---is selected. The postselection on the dark outcome involves no photon scattering, so the motional state entangled with $\ket{\downarrow}$ is not disturbed by the photon recoil. We refer to this as parity-based filtering. The filtering is nondestructive, allowing the resulting state to be reused in subsequent operations within a single experimental run.

From Eq.~\eqref{eq:V-operator}, the filtered motional state after postselection onto \(\ket{\downarrow}\) is
\begin{equation} \label{eq:motion-projection}
\begin{split}
    & \bra{\downarrow} \hat V (\boldsymbol{\theta}_f, \phi_f) \ket{\downarrow} \ket{\psi_\mathrm{mot}} \\
    &= \cos \left( \frac{\boldsymbol{\theta}_f \cdot \hat {\boldsymbol{n}} - \phi_f }{2} \right) \ket{\psi_\mathrm{mot}},
\end{split}
\end{equation}
without normalization, where \(\ket{\psi_\mathrm{mot}}\) is the motional state. For the single-mode setting \(\boldsymbol{\theta}_f=(\pi, 0)\) and an initial coherent state \(\ket{\psi_\mathrm{sm}} = \ket{\alpha}\), the filtered state becomes
\begin{equation}
\begin{cases}
    \mathcal{N}_{\mathrm{sm}, \alpha}^+ (\ket{i\alpha} + \ket{-i\alpha}), &\text{if } \phi_f=0, \\
    \mathcal{N}_{\mathrm{sm}, \alpha}^-(\ket{i\alpha} - \ket{-i\alpha}), &\text{if } \phi_f=\pi,
\end{cases}
\end{equation}
up to a global phase, where
\(\mathcal{N}_{\mathrm{sm},\alpha}^{\pm}
=1/\sqrt{2\pm 2e^{-2|\alpha|^2}}\) and the corresponding success probabilities are
\((1\pm e^{-2|\alpha|^2})/2\).
For the two-mode setting \(\boldsymbol{\theta}_f=(\pi, \pi)\) and an initial two-mode coherent state \(\ket{\psi_\mathrm{tm}} = \ket{\alpha, \alpha}\), the filtered state becomes
\begin{equation}
\begin{cases}
    \mathcal{N}_{\mathrm{tm}, \alpha}^+ (\ket{i\alpha, i\alpha} + \ket{-i\alpha, -i\alpha}), &\text{if } \phi_f=0, \\
    \mathcal{N}_{\mathrm{tm}, \alpha}^-(\ket{i\alpha, i\alpha} - \ket{-i\alpha, -i\alpha}), &\text{if } \phi_f=\pi,
\end{cases}
\end{equation}
up to a global phase, where
\(\mathcal{N}_{\mathrm{tm},\alpha}^{\pm}=1/\sqrt{2\pm2e^{-4|\alpha|^2}}\) and the corresponding success probabilities are
\((1\pm e^{-4|\alpha|^2})/2\). The filtering operation also rotates the motional state by \(\pi/2\) counterclockwise in phase space for each filtered mode, corresponding to \(e^{i\hat n\pi/2}\). This rotation can be canceled, if necessary, by applying \(\operatorname{SDR}(-\boldsymbol{\theta}_f)\) after postselection onto \(\ket{\downarrow}\).

\begin{figure}[t]
\includegraphics[width=\columnwidth]{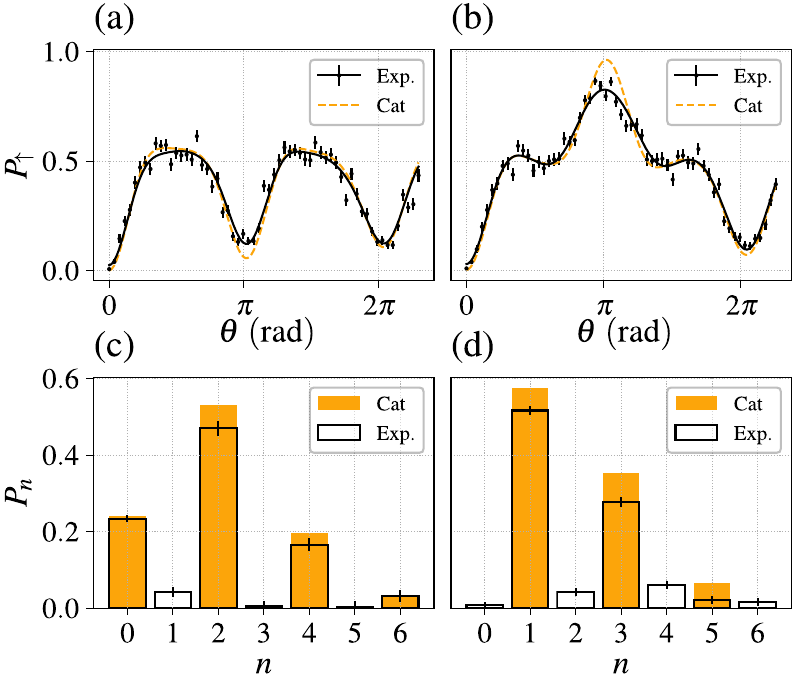}
\caption{\label{fig_singlemode_fit}
Fock-state population fitting for the single-mode case. The coherent state $\ket{\alpha}$ is filtered based on the single-mode parity. (a), (b) Time evolutions of the Ramsey sequence for even- and odd-parity-selected states, respectively, where $\theta=\chi_{\mathrm{eff},1}\sum_k t^{(k)}$. The markers denote experimental data, the solid curves are fits, and the orange dashed curves show the expected curves for ideal cat states. (c), (d) Corresponding Fock-state populations extracted from the fits (black open rectangles). The orange bars show the Fock-state distributions of the best-fit even and odd cat states, respectively.
}
\end{figure}

\begin{figure}[t]
\centerline{\includegraphics[width=\columnwidth]{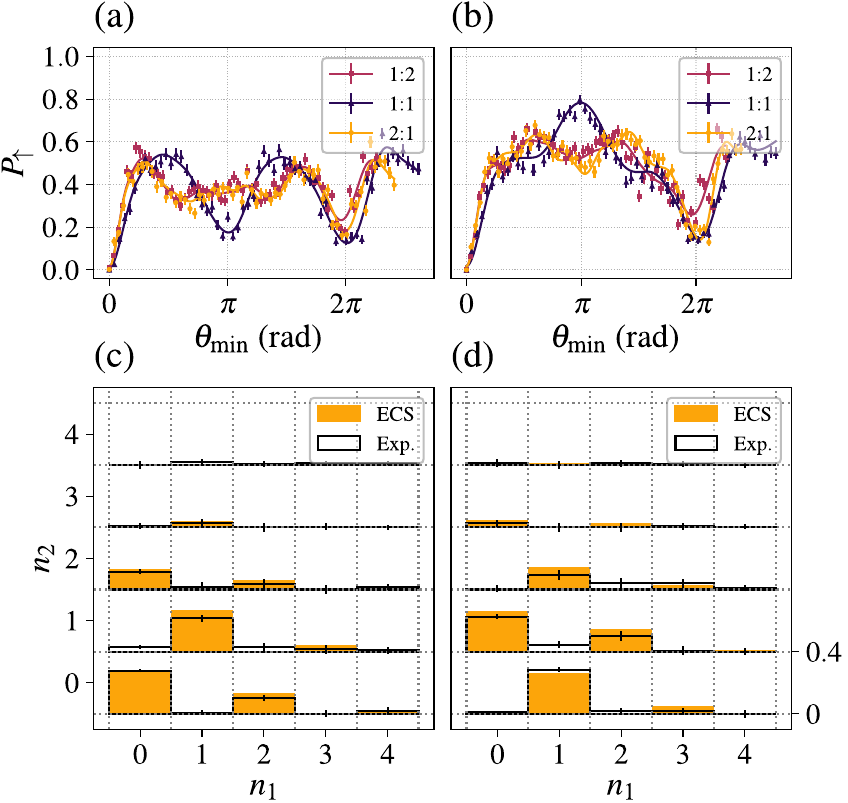}}
\caption{\label{fig_twomode_fit}
Fock-state population fitting for the two-mode case. The two-mode coherent state $\ket{\alpha,\alpha}$ is filtered based on the joint parity. (a), (b) Time evolutions of the Ramsey sequence for even- and odd-joint-parity-selected states, respectively, where $\theta_\mathrm{min}=\mathrm{min}(\chi_{\mathrm{eff},1}, \chi_{\mathrm{eff},2})\sum_k t^{(k)}$. The markers denote experimental data and the solid curves are fits. The legend labels indicate the ratio $\chi_{\mathrm{eff},1}:\chi_{\mathrm{eff},2}$.
(c), (d) Corresponding two-mode Fock-state populations extracted from fits to the measured time evolution data, shown as a $5\times5$ grid of bar plots (black open rectangles). The orange bars show the best-fit even- and odd-ECS distributions, respectively. The population axis in each cell is scaled to 0.4.
}
\end{figure}

\begin{figure}
    \centering
    \includegraphics[width=\columnwidth]{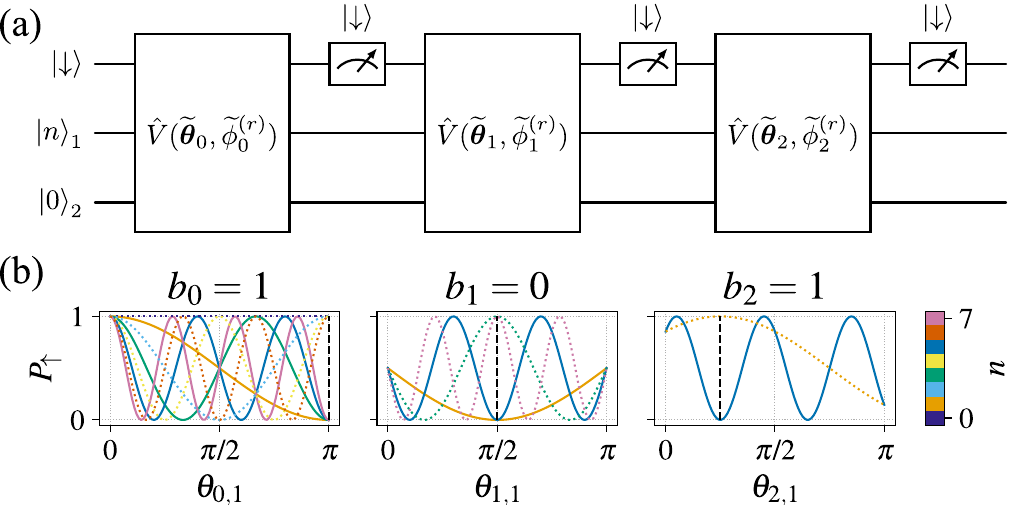} 
    \caption{\label{fig:modular-measurement-circuit}
    (a) Quantum circuit for single-shot phonon-number superparity measurement for \(m=3\). 
    The sequence alternates dispersive-shift operations 
    \(\hat{V}(\widetilde{\boldsymbol{\theta}}_\ell, \widetilde{\phi}_\ell^{(r)})\) with mid-circuit spin measurements. 
    The specific choices of rotation angles \(\widetilde{\boldsymbol{\theta}}_\ell=(\widetilde{\theta}_{\ell, 1},0)\) and target-dependent phases \(\widetilde{\phi}_\ell^{(r)}\) are given in Eq.~\eqref{eq:repeated-filter-parameter}. 
    (b) Graphical illustration of the sequential filtering process for \(m=3\) and target residue \(r=5\), for which \(\mathbf b_3(5)=(1,0,1)\).
    At each step \(\ell\), we plot the spin population \(P_\uparrow\) as a function of the rotation angle \(\theta_{\ell,1}\in[0,\pi]\) in \(\hat V((\theta_{\ell,1},0),\widetilde{\phi}_\ell^{(r)})\), while the chosen filtering angles \(\widetilde{\theta}_{\ell',1}=\pi/2^{\ell'}\) are applied in all previous steps \(\ell'<\ell\).
    The \(P_\uparrow\) oscillations are plotted for different Fock states \(\ket n\), with colors defined on the right. 
    The vertical dashed lines indicate the chosen filtering angles \(\widetilde{\theta}_{\ell, 1}=\pi/2^{\ell}\), at which each curve gives either \(P_\uparrow=0\) or \(P_\uparrow=1\). 
    We postselect the \(P_\uparrow=0\) components, shown as solid curves, and discard the \(P_\uparrow=1\) components, shown as dotted curves. 
    The discarded components are therefore absent from the next step. 
    After the final step, only \(\ket{5}\) remains among the plotted \(n=0,\ldots,7\) states, indicating successful filtering onto the target residue subspace \(\mathcal H_3^{(5)}\).}
\end{figure}

\begin{figure*}[t]
\centerline{\includegraphics[width=\textwidth]{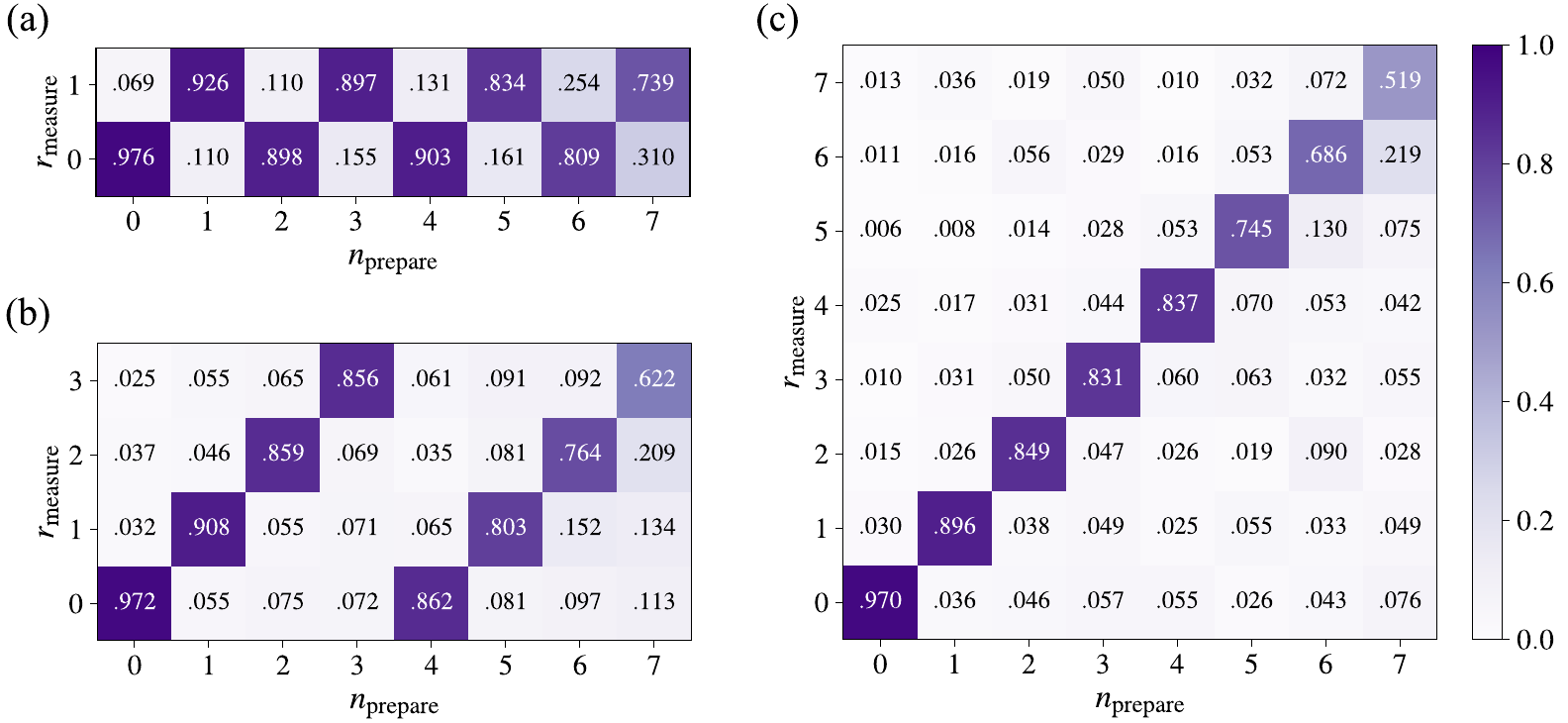}} 
\caption{\label{fig4} Experimental readout of the target residue-subspace population inferred from the mid-circuit measurement outcomes. Here, \(n_\mathrm{prepare}\) denotes the prepared Fock state, and \(r_\mathrm{measure}\) labels the target residue subspace \(\mathcal{H}_{m}^{(r_\mathrm{measure})}\), with \(m=1\) in (a), \(m=2\) in (b), and \(m=3\) in (c). The inferred populations therefore quantify the superparity measurement fidelity for the prepared Fock states. Each cell in (c) is obtained from 500 repetitions of the full measurement sequence, and the data shown in (a)--(c) are derived from the same dataset. For each given $m$, the population is inferred from the spin outcomes obtained through the first $m$ filtering steps; consequently, data with \(r_\mathrm{measure}\) values belonging to the same residue class modulo \(2^m\) are indistinguishable and are therefore combined. 
Accordingly, each cell in the panels of given $m$ contains \(500\times 2^{3-m}\) shots. }
\end{figure*}

We generate nonclassical motional states by filtering and then extract their Fock-state distributions to demonstrate the performance of both operations as shown in Fig.~\ref{fig1}(c).
We fit the data using a model that accounts for the nonlinearity of the interaction. The explicit form of the fitting function is given in Methods.
In the single-mode case, we prepare the initial state $\ket{\downarrow}\ket{\alpha}$ with a target amplitude $\alpha = 1.5$. We postselect even or odd parity components of the coherent state $\ket{\alpha}$, and observe time evolutions of the $\ket{\uparrow}$ state population. We first fit the time evolutions to extract the Fock-state populations. We then compare the results with the even and odd cat states, $\ket{\alpha} \pm \ket{-\alpha}$, where the fitted values of $\alpha$ are $1.46(1)$ and $1.40(1)$, respectively. The measurement and fitting results are shown in Fig.~\ref{fig_singlemode_fit}. We obtain the single-mode parity expectation values $\langle{e^{i\pi \hat n_1}}\rangle$ of $0.90(5)$ and $-0.73(4)$ for the even and odd parity-selected states, respectively.

In the two-mode case, we prepare the initial state $\ket{\downarrow}\ket{\alpha_1,\alpha_2}$ with target amplitudes $\alpha_1 = \alpha_2 = 1.0$. We then postselect the even- or odd-parity components of the two-mode coherent state $\ket{\alpha_1, \alpha_2}$ using the joint-parity filter. To extract the two-mode Fock-state populations, a single time evolution at one rational value of $\chi_{\mathrm{eff},1}/\chi_{\mathrm{eff},2}$ is generally insufficient because of degeneracies of the dispersive shift. We therefore acquire data at multiple rational ratios, $\chi_{\mathrm{eff},1}/\chi_{\mathrm{eff},2} \in \{0.5,\,1,\,2\}$, and jointly fit the corresponding time evolutions to obtain the two-mode Fock-state populations for the even- and odd-joint-parity-selected states, as shown in Fig.~\ref{fig_twomode_fit}. We compare the fitted distributions with those of even and odd ECSs, $\ket{\alpha,\alpha} \pm \ket{-\alpha, -\alpha}$, with fitted amplitudes $\alpha = 0.99(1)$ and $1.03(2)$, respectively. We obtain joint-parity expectation values $\langle{e^{i\pi (\hat n_1+\hat n_2)}}\rangle$ of $0.73(6)$ and $-0.65(7)$ for the even- and odd-joint-parity-selected states, respectively.

In both single- and two-mode experiments, the time evolution of the spin population does not fully return to $P_\uparrow=0$ as the Lamb-Dicke approximation breaks down at higher phonon numbers~\cite{leibfriedQuantumDynamicsSingle2003, vogelNonlinearJaynesCummingsDynamics1995, chengNonlinearQuantumRabi2018a}. 
To account for this effect, we derive and fit a model that includes the nonlinearity of the dispersive shift at large Fock numbers. This nonlinearity keeps $P_\uparrow$ in Figs.~\ref{fig_singlemode_fit} and \ref{fig_twomode_fit} from reaching zero at $\theta = 2\pi$ (See Methods).

\subsection{Single-shot phonon-number superparity measurement}

We realize a single-shot nondestructive measurement of phonon number modulo \(2^m\), which we refer to as an \(m\)-bit superparity measurement~\cite{dengQuantumenhancedMetrologyLarge2024, wangEfficientMultiphotonSampling2020, liuHybridOscillatorQubitQuantum2024}.
This measurement is enabled by the linear dependence of the dispersive shift on phonon number, which makes the $\operatorname{SDR}$ phases for different \(n\) align periodically at selected evolution times.

For a non-negative integer \(x\), we define \(b_j(x)\in\{0,1\}\) as the \(j\)-th binary digit of \(x\), with \(b_0(x)\) being the least significant bit.
Equivalently,
\begin{equation}
    x=\sum_{j=0}^{\infty} b_j(x)2^j .
\end{equation}
For \(s\ge 0\), we define the lower \(s\)-bit string of \(x\) as
\begin{equation}
    \mathbf b_s(x)
    \coloneqq
    \left(
    b_0(x), b_1(x), \dots, b_{s-1}(x)
    \right),
\end{equation}
where \(\mathbf b_0(x)\) denotes the empty string.
For an \(m\)-bit superparity measurement, the target residue is denoted by
\(r\in\{0,\dots,2^m-1\}\).
The corresponding residue subspace is
\begin{equation}
    \mathcal{H}_{m}^{(r)}
    \coloneqq
    \operatorname{span}
    \left\{
    \ket n:
    \mathbf b_m(n)=\mathbf b_m(r)
    \right\}.
    \label{eq:residue-subspace}
\end{equation}
Equivalently, \(\mathcal{H}_{m}^{(r)}\) is the subspace spanned by Fock states satisfying
\(n\equiv r \pmod{2^m}\).

To project the motional state onto the target residue subspace
\(\mathcal{H}_{m}^{(r)}\), we sequentially apply the filtering operations
\(\hat{V}(\widetilde{\boldsymbol{\theta}}_\ell,\widetilde{\phi}_\ell^{(r)})\), each followed by
postselection onto \(\ket{\downarrow}\), as illustrated in
Fig.~\ref{fig:modular-measurement-circuit}(a).
At step \(\ell=0,\dots,m-1\), we choose
\begin{equation}
    \widetilde{\boldsymbol{\theta}}_\ell
    =
    (\widetilde{\theta}_{\ell,1},0)
    =
    \left(\frac{\pi}{2^{\ell}},0\right),
    \quad
    \widetilde{\phi}_\ell^{(r)}
    =
    \frac{\pi}{2^{\ell}}
    \sum_{j=0}^{\ell} b_j(r)2^j .
\label{eq:repeated-filter-parameter}
\end{equation}
Conditioned on successful nondestructive postselection onto the dark
\(\ket{\downarrow}\) state, step \(\ell\) filters the motional state from
\(\mathcal H_{\ell}^{(r)}\) into \(\mathcal H_{\ell+1}^{(r)}\) by checking
whether \(b_\ell(n)\) agrees with the target bit \(b_\ell(r)\).
Specifically, as shown in Eq.~\eqref{eq:motion-projection}, the motional
state is filtered by the non-unitary operator
\begin{equation}
    \hat F_\ell^{(r)}
    \coloneqq
    \cos\left(
    \frac{\widetilde{\theta}_{\ell,1}\hat n_1-\widetilde{\phi}_\ell^{(r)}}{2}
    \right).
\end{equation}
For a Fock-state component \(\ket n \in \mathcal H_{\ell}^{(r)}\), namely
\(
    \mathbf b_\ell(n)=\mathbf b_\ell(r),
\)
the action of filter \(\ell\) is
\begin{equation}
\hat F_\ell^{(r)}\ket n
=
\begin{cases}
    (-1)^{b_{\ell+1}(n)}\ket n,
    & b_\ell(n)=b_\ell(r),\\
    0,
    & b_\ell(n)\neq b_\ell(r).
\end{cases}
\label{eq:filter-action}
\end{equation}
Thus, step \(\ell\) accepts only the components whose
binary digit \(b_\ell(n)\) agrees with the target bit \(b_\ell(r)\).
After successful postselection through steps \(0,\dots,\ell\), the surviving
components satisfy
\(
    \mathbf b_{\ell+1}(n)=\mathbf b_{\ell+1}(r).
\)
Repeating this procedure for \(\ell=0,\dots,m-1\) yields
\(\mathbf b_m(n)=\mathbf b_m(r)\), thereby projecting the motional state onto
\(\mathcal{H}_{m}^{(r)}\), up to number-dependent phase factors.

Although \(\hat F_\ell^{(r)}\) preserves the Fock-basis populations of the
accepted components up to normalization, it is not an exact projection
operator because it is not idempotent,
\begin{equation}
    \left(\hat F_\ell^{(r)}\right)^2
    \neq
    \hat F_\ell^{(r)} .
\end{equation}
For \(\ell<m-1\), any state that survives all \(m\) filtering steps satisfies
\(
    b_{\ell+1}(n)=b_{\ell+1}(r).
\)
Hence the phase factor \((-1)^{b_{\ell+1}(n)}\) acquired at step \(\ell\) is
a global phase within the final residue subspace \(\mathcal{H}_{m}^{(r)}\).
In contrast, the final step \(\ell=m-1\) leaves the factor
\((-1)^{b_m(n)}\), which is not fixed within \(\mathcal{H}_{m}^{(r)}\).
This residual number-dependent phase can be removed by applying the
compensating rotation \(\operatorname{SDR}(-\pi/2^{m-1},0)\) after
postselection.
Consequently, up to an irrelevant global phase, the projection operator onto
\(\mathcal{H}_{m}^{(r)}\) can be written as
\begin{equation}
    \hat \Pi_{m}^{(r)}
    =
    \operatorname{SDR}
    \left(-\frac{\pi}{2^{m-1}},0\right)
    \prod_{\ell=0}^{m-1}
    \hat F_\ell^{(r)} .
    \label{eq:total-projection-operator}
\end{equation}

In Fig.~\ref{fig:modular-measurement-circuit}(b), we illustrate the filtering
process for \(m=3\) and \(r=5\) as an example, for which
\(
    \mathbf b_3(5)=(1,0,1).
\)
The first step, \(\ell=0\), reduces to an ordinary parity measurement.
Since \(b_0(5)=1\), the operation \(\hat{V}(\widetilde{\boldsymbol{\theta}}_0,\widetilde{\phi}_0^{(5)})\) maps odd Fock states onto the spin state \(\ket{\downarrow}\), while mapping even Fock states onto \(\ket{\uparrow}\).
Postselection onto the dark state \(\ket{\downarrow}\) therefore selects the
components with the desired least significant bit.
For \(b_0(r)=0\), in contrast, the phase offset is \(\widetilde{\phi}_0^{(r)}=0\) rather than \(\pi\), so the \(P_\uparrow\) oscillation starts from 0. This produces the inverted trace shown in the \(b_0\) panel of Fig.~\ref{fig:modular-measurement-circuit}(b).

In the next step, \(\ell=1\), only odd Fock states remain after the first postselection.
The phase offset \(\widetilde{\phi}_1^{(5)}=\pi/2\) makes the \(P_\uparrow\) oscillation start from \(0.5\).
At the filtering condition \(\widetilde{\theta}_{1,1}=\pi/2\), these states are separated into two groups according to \(b_1(n)=0\) or \(1\), which are mapped onto \(\ket{\downarrow}\) and \(\ket{\uparrow}\), respectively.
In the final step, \(\ell=2\), the remaining Fock states are again separated into two groups according to \(b_2(n)\) at the filtering condition \(\widetilde{\theta}_{2,1}=\pi/4\).
This example shows that the phase offset \(\widetilde{\phi}_\ell^{(r)}\) plays an essential role in partitioning the bosonic subspace and assigning the two resulting subspaces to the spin states \(\ket{\downarrow}\) and \(\ket{\uparrow}\).

Using this protocol, one can also estimate the population
\(p_{m}^{(r)}(\hat{\rho}_\mathrm{mot})\) of an initial motional state
\(\hat{\rho}_\mathrm{mot}\) in the target residue subspace
\(\mathcal{H}_{m}^{(r)}\), defined as
\begin{equation}
p_{m}^{(r)}(\hat{\rho}_\mathrm{mot})
\coloneqq
\operatorname{tr}\!\left(
\hat{\rho}_\mathrm{mot}
\hat{\Pi}_{m}^{(r)}
\right).
\end{equation}
Ideally, this population is the joint probability to pass all the filters
for the target residue \(r\), given by the product of the conditional
filter-pass probabilities over steps \(0,\dots,m-1\),
\begin{equation}
    p_{m}^{(r)}(\hat{\rho}_\mathrm{mot})
    =
    \prod_{\ell=0}^{m-1}
    P_{\downarrow}^{(\ell)} .
    \label{eq:subspace-population}
\end{equation}
Here, \(P_{\downarrow}^{(\ell)}\) is the conditional probability of obtaining
the spin outcome \(\ket{\downarrow}\) at step \(\ell\), given that the
outcomes at all previous steps \(0,\dots,\ell-1\) were also
\(\ket{\downarrow}\).
See Methods for more details of the estimation protocol.
For this population-estimation scheme, it is not necessary to compensate for
the bosonic phase rotation as in Eq.~\eqref{eq:total-projection-operator},
because only the diagonal elements of \(\hat{\rho}_\mathrm{mot}\) contribute
to \(p_{m}^{(r)}(\hat{\rho}_\mathrm{mot})\).

To demonstrate this protocol, we prepare Fock states
\(\ket{n_{\mathrm{prepare}}}\) in mode 1 with
\(n_{\mathrm{prepare}}=0,\dots,7\), and then apply the sequential filtering
operations for \(m=3\), targeting the subspace
\(\mathcal{H}_{m}^{(r_\mathrm{measure})}\).
We then estimate
\(p_{m}^{(r_\mathrm{measure})}(\hat{\rho}_\mathrm{prepare})\)
from the measured conditional spin-\(\ket{\downarrow}\) populations at each
step, where
\(
    \hat{\rho}_\mathrm{prepare}
    =
    \ket{n_{\mathrm{prepare}}}
    \bra{n_{\mathrm{prepare}}}.
\)
Because \(\ket{n_\mathrm{prepare}}\) belongs to the residue subspace specified
by \(\mathbf b_m(n_\mathrm{prepare})\), the expected population is
\begin{equation}
\label{eq:ideal-subspace-population}
p_{m}^{(r_\mathrm{measure})}
(\hat{\rho}_\mathrm{prepare})
=
\begin{cases}
    1,
    &
    \mathbf b_m(n_\mathrm{prepare})
    =
    \mathbf b_m(r_\mathrm{measure}),
    \\
    0,
    &
    \mathrm{otherwise}.
\end{cases}
\end{equation}
Equivalently, the expected population is unity when
\(n_\mathrm{prepare}\equiv r_\mathrm{measure}\pmod{2^m}\) and zero otherwise.
The measured populations for \(m=3\) are shown in Fig.~\ref{fig4}(c), showing
strong diagonal contrast with suppressed off-diagonal populations.

We also infer the populations for \(m=1\) and \(m=2\) from the same dataset,
as shown in Figs.~\ref{fig4}(a) and (b), respectively, by using only the first
\(m\) spin outcomes from each filtering sequence.
Since these outcomes only determine the lower \(m\)-bit string
\(\mathbf b_m(n)\), data corresponding to \(r_\mathrm{measure}\) values with
the same lower \(m\)-bit string---for example, \(0\) and \(4\) when
\(m=2\)---are combined.
The results are consistent with the prediction of
Eq.~\eqref{eq:ideal-subspace-population}.
The agreement with the ideal values therefore provides a measure of the
superparity measurement fidelity for the prepared Fock states.
The measured infidelity increases with the prepared phonon number,
probably due
to the larger $\operatorname{SDR}$ decay rate induced by beam-intensity noise.

\section{Discussion}

We present a systematic study of multimode dispersive shifts in a trapped-ion system and establish a multimode $\operatorname{SDR}$ Ramsey sequence that enables two complementary phonon-number measurement schemes: multimode phonon-number measurement and superparity measurement. Beyond population estimation, this approach can be extended to density matrix reconstruction by combining displacements with phonon-number measurements~\cite{leibfriedExperimentalDeterminationMotional1996,jiaDeterminationMultimodeMotional2022}.
The second scheme, phonon-number superparity measurements, highlights the role of $\operatorname{SDR}$ in bosonic quantum error correction. Our nondestructive single-shot modulo $2^m$ measurements can serve as syndrome measurements for rotation-symmetric bosonic codes, including cat and binomial codes~\cite{michaelNewClassQuantum2016,ofekExtendingLifetimeQuantum2016,grimsmoQuantumComputingRotationSymmetric2020}. 
Although the present implementation targets a preselected residue subspace, the scheme can be further generalized to resolve the full superparity bit string by incorporating beam-splitter operations~\cite{houCoherentCouplingNondestructive2024}.
More broadly, multimode $\operatorname{SDR}$ also implements controlled parity operators. In trapped-ion experiments, parity information has so far been inferred indirectly from characteristic functions~\cite{fluhmannDirectCharacteristicFunctionTomography2020} or reconstructed from phonon-number distributions, while single-shot parity measurements have also been demonstrated~\cite{ganHybridQuantumComputing2020,jeonMultimodeBosonicState2025}.
The present protocol can serve as a single-shot multimode (including single-mode) parity measurement for multimode Wigner function measurement or parity-based quantum metrology~\cite{gerryParityOperatorQuantum2010,birrittellaParityOperatorApplications2021}. 
Beyond measurement, the underlying $\operatorname{SDR}$ operation can be used as a controlled-parity unitary for bosonic quantum error-correction codes. 
Single-mode $\operatorname{SDR}$ can serve as a spin-controlled single-qubit gate~\cite{grimsmoQuantumComputingRotationSymmetric2020}, and two-mode $\operatorname{SDR}$ can serve as a joint-$4$ parity operator to perform a two-qubit gate for certain bosonic codes~\cite{tsunodaErrorDetectableBosonicEntangling2023} or as a spin-controlled number-difference operator related to pair-coherent states and pair-cat states~\cite{gertlerExperimentalRealizationCharacterization2023,albertPaircatCodesAutonomous2019}. Moreover, stabilization of the dispersive shift and combining it with multi-tone microwave control could enable a SNAP gate for bosonic state control~\cite{heeresImplementingUniversalGate2017, krastanovUniversalControlOscillator2015,eickbuschFastUniversalControl2022}.

\section{methods}
\subsection{Bosonic state preparation}
We prepare both Fock states and coherent states from the initial state $\ket{\downarrow}\ket{0,0}$, obtained by optical pumping of the spin state and sideband cooling of the two motional modes.

\emph{Fock-state preparation.}
To prepare a two-mode Fock state $\ket{n_1,n_2}$, we apply blue-sideband and red-sideband $\pi$ pulses alternately that add phonons one by one. We first prepare mode~1 to obtain $\ket{n_1,0}$ and then apply the same procedure to mode~2 to reach the target state $\ket{n_1,n_2}$.

\emph{Coherent-state preparation.}
To prepare a two-mode coherent state $\ket{\alpha_1,\alpha_2}$, we first apply a carrier $\pi/2$ pulse to initialize the spin in the superposition state
$\ket{+}=(\ket{\downarrow}+\ket{\uparrow})/\sqrt{2}$.
We then apply spin-dependent forces sequentially to the two motional modes. Then we apply another $\pi/2$ pulse to obtain the final state $\ket{\downarrow}\ket{\alpha_1,\alpha_2}$.

\subsection{Determination of phonon-number distribution beyond the Lamb-Dicke approximation}

To describe the Ramsey signal at larger phonon numbers appropriately, we fitted the data using a model that retains the number dependence of the dispersive shifts beyond the Lamb-Dicke approximation~\cite{vogelNonlinearJaynesCummingsDynamics1995}.

For a single off-resonant red- or blue-sideband drive, the nonlinear spin-motion interaction can be written as
\begin{equation}
\hat H_I(t)\simeq
\hbar\sum_j \frac{\eta_j\Omega}{2}
\Big[
\hat{\sigma}_\pm \tilde a_j\,e^{\mp i\delta_j t}
+
\hat{\sigma}_\mp \tilde a_j^\dagger\,e^{\pm i\delta_j t}
\Big],
\end{equation}
with
\begin{equation}
\tilde a_j=f_j(\hat n_j)\hat a_j \hat M_j,
\qquad
f_j(\hat n_j)=e^{-\eta_j^2/2}\frac{L_{\hat n_j}^{(1)}(\eta_j^2)}{\hat n_j+1},
\end{equation}
\begin{equation}
\hat M_j=\prod_{\ell\neq j} e^{-\eta_\ell^2/2}L_{\hat n_\ell}(\eta_\ell^2).
\end{equation}
Here \(L_n\) and \(L_n^{(1)}\) are Laguerre and associated Laguerre polynomials. Since \(f_j(\hat n_j)\) and \(\hat M_j\) are diagonal in the multimode Fock basis, we denote their eigenvalues on \(\ket{\boldsymbol n}\) by \(f_j(n_j)\) and \(M_j(\boldsymbol n)\), respectively.

In the dispersive regime, the corresponding nonlinear effective Hamiltonian is
\begin{equation}
\hat H_{\mathrm{disp}}^{(\mathrm{nl})}
=
\hbar\sum_j \frac{\chi_j}{2}\hat\sigma_z\,\hat M_j^2
\left[
f_j(\hat n_j)^2(\hat n_j+1)
+
f_j(\hat n_j-1)^2\hat n_j
\right],
\end{equation}
with $\chi_j=-\frac{\eta_j^2\Omega^2}{4\delta_j}$. For an initial motional state \(\hat\rho_{\mathrm{mot}}\), the Ramsey signal is then
\begin{equation}
P_{\uparrow}(t)
=
\frac{1}{2}
-\frac{1}{2}
\sum_{\boldsymbol n}
p_{\boldsymbol n}
\cos\!\left[\Phi_{\boldsymbol n}(t)-\phi_{\mathrm{off}}(t)\right],
\label{eq:methods_ramsey_nonlinear}
\end{equation}
where \(p_{\boldsymbol n}=\langle \boldsymbol n|\hat\rho_{\mathrm{mot}}|\boldsymbol n\rangle\), \(\sum_{\boldsymbol n}p_{\boldsymbol n}=1\), and
\begin{equation}
\Phi_{\boldsymbol n}(t)
=
t\sum_j \chi_j\,M_j(\boldsymbol n)^2
\left[
f_j(n_j)^2(n_j+1)
+
f_j(n_j-1)^2 n_j
\right].
\end{equation}
In the Lamb-Dicke limit, \(f_j(n_j)\to 1\) and \(M_j(\boldsymbol n)\to 1\), so that Eq.~\eqref{eq:methods_ramsey_nonlinear} reduces to the linear fitting model Eq.~\eqref{eq:Pr-up-cos}. We use Eq.~\eqref{eq:methods_ramsey_nonlinear} to fit the experimental data in Figs.~\ref{fig_singlemode_fit} and ~\ref{fig_twomode_fit}. A detailed derivation is provided in the Supplementary Information.

\subsection{Residual shift offset compensation}
We employ a selective decoupling scheme to cancel out the phase accumulation by the carrier-induced AC-Stark shift. However, even when the contribution of the carrier transition is nullified, there is a remaining offset $\chi_{\mathrm{eff},j}$ in addition to the desired phonon-number-dependent shift $2\chi_{\mathrm{eff},j} n_j$. To cancel this offset, we shift the phase of the second microwave $\pi/2$ pulse when implementing the Ramsey sequence of $\hat V(\boldsymbol{\theta}, \phi)$ by $\phi_\mathrm{off} = \tilde\Delta_\mathrm{off} t$, where $\tilde\Delta_\mathrm{off}$ denotes the \textit{effective} residual shift for compensation and $t=t^{(1)}+t^{(2)}$. The term \textit{effective} implies that $\tilde\Delta_\mathrm{off}$ is defined by the accumulated total phase divided by the total interaction time $t$. This definition simplifies the calibration, although the actual residual shifts may be different in step segments $(1)$ and $(2)$. Ideally, this compensation results in a flat signal for the motional vacuum state. However, in experiment, a contrast decay is observed probably due to motional decoherence, beam intensity fluctuations, and a drifting repetition rate of the pulse laser. Therefore, we design a calibration procedure as follows:
\begin{enumerate}
    \item Prepare the state $\ket\downarrow\ket{0,0}$.
    \item Apply the Ramsey sequence $\hat R_{\Delta_\mathrm{off}t_\mathrm{cal}}(-\frac \pi 2)\hat{U}^{(2)}\hat{R}_y(\pi)\,\hat{U}^{(1)}\hat R_x(\frac \pi 2)$ for a fixed total interaction time $t_\mathrm{cal}$ and measure the spin.
    \item Repeat the previous steps with different $\Delta_\mathrm{off}$ values and determine $\tilde\Delta_\mathrm{off}$ by finding the center of a sinusoidal dip by fitting.
\end{enumerate}
A longer interaction time $t_\mathrm{cal}$ provides higher fitting resolution, unless it is longer than the coherence time of the signal. We empirically choose $t_\mathrm{cal}$ to be about $4~\mathrm{ms}$. After a calibration, we store the calibrated $\tilde\Delta_\mathrm{off}$ value for subsequent experiments.

Before every experimental set---once for a certain parameter set, not for every shot---we check if the residual shift is sufficiently low, so that the signal is flat enough. The criterion is $P_\uparrow < 0.1$ at $t_\mathrm{cal}$ with the stored $\tilde\Delta_\mathrm{off}$. If the test fails, we recalibrate $\tilde\Delta_\mathrm{off}$ and update the stored value.

\subsection{Parity-based filtering condition}
To implement the parity operator for the parity-based filtering experiments, we calibrate the interaction time $t_\pi$ to achieve $\boldsymbol{\theta}=(\pi, 0)$---or $(\pi, \pi)$ in two-mode cases---in $\hat V(\boldsymbol{\theta}, \phi)$. For an even-parity state, the $\ket{\uparrow}$ state population at $t_\pi$ under $\hat V(\boldsymbol{\theta}, 0)$ is ideally zero. The calibration procedure is similar to that of the residual offset compensation. We simply prepare the motional Fock state $\ket{2, 0}$ and observe the time evolution of $P_\uparrow$ under $\hat V(\boldsymbol{\theta}, \phi)$. Then we find the center of the first sinusoidal dip by fitting. This works for both single- and two-mode parities as both are even. We calibrate $t_\pi$ before every experiment set.

\subsection{Estimation of superparity population}

As illustrated in Fig.~\ref{fig4}(a), the single-shot phonon-number superparity measurement consists of a sequence of filtering steps.
To measure the population in the target subspace \(\mathcal{H}_{m}^{(r_\mathrm{measure})}\), we choose the filter parameters for step \(\ell=0,\dots,m-1\) according to Eq.~\eqref{eq:repeated-filter-parameter}, with the target residue \(r=r_\mathrm{measure}\).
The target subspace is spanned by all Fock states satisfying
\(
    \mathbf b_m(n)=\mathbf b_m(r_\mathrm{measure}),
\)
or equivalently,
\(
    n \equiv r_{\mathrm{measure}} \pmod{2^m}.
\)

\paragraph{State-discrimination threshold for postselection.}
In the experiment, the spin state is inferred using a photon-count threshold.
If the number of detected photons \(N\) exceeds a threshold \(N_t\), the outcome is assigned to the bright state \(\ket{\uparrow}\).
Otherwise, for \(N\le N_t\), it is assigned to the dark state \(\ket{\downarrow}\).
In our system, using \(N_t=1\), the typical state-discrimination fidelities are
\(P(\uparrow|\uparrow)=0.95\text{--}0.97\) and
\(P(\uparrow|\downarrow)=0.00\text{--}0.01\).
The dominant detection error is therefore
\(P(\downarrow|\uparrow)=0.03\text{--}0.05\), which can lead to false acceptance during postselection and hence reduce the fidelity of the heralded projection.

To suppress this error, we use a stricter threshold \(N_t=0\) for the pass/reject decision in the postselection step, discarding all experimental runs with nonzero photon counts.
In addition, when estimating the superparity population, we compensate for the residual detection error \(P(\downarrow|\uparrow)\) by rescaling the measured spin populations.
The correction procedure is described in detail in Appendix~D.2 of Ref.~\cite{jeonMultimodeBosonicState2025}.

\paragraph{Target-subspace population estimation.}
We now describe how the population in the target modular subspace is estimated from repeated experimental realizations.
Although the stricter threshold \(N_t=0\) improves the quality of the heralding condition, it lowers the state-discrimination fidelity.
We therefore use \(N_t=0\) for the filter pass/reject decision and \(N_t=1\) for spin-state discrimination.

To distinguish these two roles, we define \(A_\ell\) and \(B_\ell\) as the events
\begin{equation}
\begin{split}
A_\ell &: N_\ell = 0,\\
B_\ell &: N_\ell \le 1,
\end{split}
\end{equation}
where \(N_\ell\) is the number of detected photons at step \(\ell\).
Thus, \(A_\ell\) represents the strict heralding condition used for postselection, whereas \(B_\ell\) represents the dark-state assignment used for estimating spin populations.
For reference, we also define \(C_\ell\) as the event that the true collapsed spin state after step \(\ell\) is \(\ket{\downarrow}\).

The quantity of interest is the population in the target subspace, as shown in Eq.~\eqref{eq:subspace-population}.
For the target residue \(r_\mathrm{measure}\), it can be written as
\begin{equation}
p_{m}^{(r_{\mathrm{measure}})}
=
P(\boldsymbol{C}_{m-1})
=
P(C_0)
\prod_{\ell=1}^{m-1}
P(C_\ell \mid \boldsymbol{C}_{\ell-1}),
\end{equation}
where
\begin{equation}
\boldsymbol{C}_\ell
\equiv
\bigcap_{j=0}^{\ell} C_j ,
\end{equation}
and \(\boldsymbol{A}_\ell\) and \(\boldsymbol{B}_\ell\) are defined analogously.

Because \(B_\ell\) approximates \(C_\ell\) well at the level of single-step spin-state discrimination, one has \(P(B_\ell)\approx P(C_\ell)\), up to the detection infidelity, which is compensated by the rescaling described above.
However, the conditional probability \(P(B_\ell\mid \boldsymbol{B}_{\ell-1})\) does not generally approximate \(P(C_\ell\mid \boldsymbol{C}_{\ell-1})\), because the ensemble selected by \(\boldsymbol{B}_{\ell-1}\) is contaminated by falsely accepted bright-state events.
Instead, we condition on the stricter postselection event \(\boldsymbol{A}_{\ell-1}\), for which false acceptance of \(\ket{\uparrow}\) is strongly suppressed.
Under this condition,
\begin{equation}
P(B_\ell \mid \boldsymbol{A}_{\ell-1})
\approx
P(C_\ell \mid \boldsymbol{C}_{\ell-1}) .
\end{equation}
Accordingly, we estimate the target-subspace population as
\begin{equation}
p_{m}^{(r_{\mathrm{measure}})}
\approx
P(B_0)
\prod_{\ell=1}^{m-1}
P(B_\ell \mid \boldsymbol{A}_{\ell-1}) .
\label{eq:p_n_estimation}
\end{equation}

\begin{acknowledgments}
\paragraph{Funding}This work has been supported by the National Research Foundation of Korea (NRF) grant (No. RS-2024-00442855, No. RS-2024-00413957, No. RS-2024-00466865), and the Institute of Information \& Communications Technology Planning \& Evaluation (IITP) grant (No. RS-2022-II221040), all of which are funded by the Korean government (MSIT).
\end{acknowledgments}

\bibliography{dispersive_shift_ref}

\end{document}


\preprint{APS/123-QED}

\title{Supplemental Material: Multimode Phonon-Number Measurement and Superparity Measurement using Dispersive Shifts in a Trapped Ion}

\author{Wonhyeong Choi}
\thanks{These two authors contributed equally}
\affiliation{Department of Computer Science and Engineering, Seoul National University}
\affiliation{Automation and Systems Research Institute, Seoul National University}
\affiliation{NextQuantum, Seoul National University}

\author{Jiyong Kang}
\thanks{These two authors contributed equally}
\affiliation{Department of Computer Science and Engineering, Seoul National University}
\affiliation{Automation and Systems Research Institute, Seoul National University}
\affiliation{NextQuantum, Seoul National University}

\author{Kyunghye Kim}
\affiliation{Department of Computer Science and Engineering, Seoul National University}
\affiliation{Automation and Systems Research Institute, Seoul National University}
\affiliation{NextQuantum, Seoul National University}

\author{Jaehun You}
\affiliation{Department of Computer Science and Engineering, Seoul National University}
\affiliation{Automation and Systems Research Institute, Seoul National University}
\affiliation{NextQuantum, Seoul National University}

\author{Kyungmin Lee}
\affiliation{Department of Computer Science and Engineering, Seoul National University}
\affiliation{Automation and Systems Research Institute, Seoul National University}
\affiliation{NextQuantum, Seoul National University}

\author{Taehyun Kim}
\email{taehyun@snu.ac.kr}
\affiliation{Department of Computer Science and Engineering, Seoul National University}
\affiliation{Automation and Systems Research Institute, Seoul National University}
\affiliation{NextQuantum, Seoul National University}
\affiliation{Institute of Applied Physics, Seoul National University}
\date{\today}
\maketitle
\onecolumngrid
\setcounter{equation}{0}
\renewcommand{\theequation}{S\arabic{equation}}
\setcounter{figure}{0}
\renewcommand{\thefigure}{S\arabic{figure}}
\setcounter{table}{0}
\renewcommand{\thetable}{S\arabic{table}}

\section{S1. Derivation of the Multimode dispersive shift}

\subsection{S1.1. Multimode linear Jaynes-Cummings model}

We consider a two-level system with states $\ket{\downarrow},\ket{\uparrow}$ and splitting $\omega_0$,
coupled to bosonic modes $j$ of frequency $\omega_j$ with ladder operators $\hat a_j$.
A laser drive of frequency $\omega_L$ and Rabi frequency $\Omega$ addresses the transition.
In the lab frame the Hamiltonian is
\begin{equation}
\hat H_{\rm lab}(t)
= \hbar\frac{\omega_0}{2}\hat{\sigma}_z
+ \hbar\sum_{j}\omega_j\,\hat a_j^\dagger \hat a_j
+ \hbar\frac{\Omega}{2}\Big[\hat{\sigma}_+ e^{i\phi}e^{-i\omega_L t}\,
\prod_{j} e^{\,i\eta_j\left(\hat{a}_j +\hat{a}_j^\dagger \right)}
+\text{h.c.}\Big],
\label{eq:H-lab-single-tone}
\end{equation}
where $\eta_j$ are the Lamb-Dicke parameters.

We define the carrier detuning $\Delta\coloneqq\omega_L-\omega_0$ and go to the interaction picture with respect to 
$\hat H_0^{\rm free}=\hbar\frac{\omega_0}{2}\hat{\sigma}_z+\hbar\sum_j \omega_j \hat a_j^\dagger \hat a_j$.
With $\hat U_0(t)=e^{-i\hat H_0^{\rm free}t/\hbar}$ we obtain
\begin{equation}
\hat H_{I}(t)
= \hbar\frac{\Omega}{2}\Big[\hat{\sigma}_+ e^{i\phi}e^{-i\Delta t}\,
\prod_{j} e^{\,i\eta_j\left(\hat{a}_j e^{-i\omega_j t}+\hat{a}_j^\dagger e^{i\omega_j t}\right)}
+\text{h.c.}\Big].
\label{eq:Single-tone}
\end{equation}

We drive either the red or the blue sideband (but not both) of a single spin coupled to bosonic modes indexed by $j$.
In the Lamb-Dicke regime we expand the displacement operator to first order, choose $\phi=-\pi/2$, and apply the rotating-wave approximation (RWA), retaining only the red-sideband (RSB) or blue-sideband (BSB) terms.
For each mode $j$ we define the detuning from its sideband by $\delta_j=\Delta-(\mp\omega_j)$, where the upper (lower) sign corresponds to the RSB (BSB).
This gives the sideband-RWA interaction Hamiltonian
\begin{equation}
\hat H_{I}(t)\simeq \hbar\sum_j g_j\Big[\hat{\sigma}_\pm \hat{a}_j\,e^{\mp i\delta_j t}
+\hat{\sigma}_\mp \hat{a}_j^\dagger\,e^{\pm i\delta_j t}\Big],
\qquad g_j\coloneqq\frac{\eta_j\Omega}{2}.
\end{equation}

It is convenient to remove the explicit time dependence by a second interaction-picture transformation generated by the detuning rotation of the motional modes.
We define the unitary change of frame by
\begin{equation}
\ket{\psi'(t)}=\hat W(t)\ket{\psi(t)},\qquad
\hat W(t)=\exp\!\Big(\mp i t\sum_j \delta_j \hat n_j\Big),
\qquad \hat n_j=\hat a_j^\dagger \hat a_j ,
\end{equation}
where the upper (lower) sign corresponds to the RSB (BSB), consistent with the sign convention in $\hat H_{I}(t)$.
With this convention, the Hamiltonian in the transformed frame is
\begin{equation}
\hat H
=\hat W\,\hat H_{I}(t)\,\hat W^\dagger
+i\hbar\,\dot{\hat W}\,\hat W^\dagger .
\end{equation}
Using $[\hat n_j,\hat a_j]=-\hat a_j$ and $[\hat n_j,\hat a_j^\dagger]=+\hat a_j^\dagger$, we obtain
\begin{equation}
\hat W\,\hat a_j\,\hat W^\dagger=\hat a_j\,e^{\pm i\delta_j t},
\qquad
\hat W\,\hat a_j^\dagger\,\hat W^\dagger=\hat a_j^\dagger\,e^{\mp i\delta_j t},
\end{equation}
so that the phase factors in $\hat H_{I}(t)$ cancel.
The second term contributes a detuning Hamiltonian,
\begin{equation}
i\hbar\,\dot{\hat W}\,\hat W^\dagger
=\pm \hbar\sum_j \delta_j \hat n_j .
\end{equation}
As a result, we obtain a time-independent form
\begin{equation}
\hat H=\hat H_0+\hat V,\qquad
\hat H_0=\pm\hbar\sum_{j}\delta_j\,\hat n_j,\qquad
\hat V=\hbar\sum_{j} g_j\big(\hat{\sigma}_{\pm} \hat{a}_j+\hat{\sigma}_{\mp} \hat{a}_j^\dagger\big),
\label{eq:model}
\end{equation}
with detunings $\delta_j$ and couplings $g_j$ for each mode $j$.
We assume the dispersive regime $|\delta_j|\gg g_j\sqrt{n_j+1}$, and we perform two steps of Schrieffer-Wolff (SW) transformations to derive Eq.~(2) in the main text~\cite{SW}.

\paragraph{First SW step.}
We choose an anti-Hermitian $\hat S_1$ that solves $[\hat S_1,\hat H_0]+\hat V=0$:
\begin{equation}
\hat S_1=\mp\sum_{j}\frac{g_j}{\delta_j}\big(\hat{\sigma}_{\pm} \hat{a}_j-\hat{\sigma}_{\mp} \hat{a}_j^\dagger\big),
\qquad \hat S_1^\dagger=-\hat S_1.
\label{eq:S1}
\end{equation}
To second order, the effective Hamiltonian is
\begin{equation}
\hat H_{\mathrm{eff}}^{(1)}=\hat H_0+\frac{1}{2}[\hat S_1,\hat V].
\end{equation}
We focus on the commutator
\begin{equation}
[\hat S_1,\hat V]
=\mp\sum_{j,k}\Big[\frac{g_j}{\delta_j}\big(\hat{\sigma}_{\pm} \hat{a}_j-\hat{\sigma}_{\mp} \hat{a}_j^\dagger\big),
\hbar g_k\big(\hat{\sigma}_{\pm} \hat{a}_k+\hat{\sigma}_{\mp} \hat{a}_k^\dagger\big)\Big].
\end{equation}
The commutator $\tfrac 12[\hat S_1,\hat V]$ consists of (\romannumeral 1) a dispersive-shift term $\hat H_{\mathrm{disp}}$ and (\romannumeral 2) a pairwise mixing term $\hat H_{\mathrm{cross}}$.

Using
$[\hat{\sigma}_{\pm} \hat{a}_j,\hat{\sigma}_{\mp}\hat{a}_j^\dagger]
=\pm\hat{\sigma}_z\!\big(\hat{n}_j+\tfrac12\big)+\tfrac12$,
the $j=k$ terms in $\tfrac 12[\hat S_1,\hat V]$ yield
\begin{equation}
\hat H_\mathrm{disp}
=\mp\frac 12\sum_{j}\Big[\frac{g_j}{\delta_j}\big(\hat{\sigma}_{\pm} \hat{a}_j-\hat{\sigma}_{\mp} \hat{a}_j^\dagger\big),
\hbar g_j\big(\hat{\sigma}_{\pm} \hat{a}_j+\hat{\sigma}_{\mp} \hat{a}_j^\dagger\big)\Big]
=\hbar\sum_{j}\,\chi_j\,\hat{\sigma}_z\!\big(\hat{n}_j+\tfrac12\big)+\text{const.},
\label{eq:H_disp}
\end{equation}
where $\chi_j=-g_j^2/\delta_j$.
This is the multimode dispersive shift in the main text, Eq.~(2).

For $j\neq k$, the commutator generates a spin-dependent beam-splitter term
\begin{equation}
\hat H_\mathrm{cross}
=\mp\frac 12\sum_{j\neq k}\Big[\frac{g_j}{\delta_j}\big(\hat{\sigma}_{\pm} \hat{a}_j-\hat{\sigma}_{\mp} \hat{a}_j^\dagger\big),
\hbar g_k\big(\hat{\sigma}_{\pm} \hat{a}_k+\hat{\sigma}_{\mp} \hat{a}_k^\dagger\big)\Big]
=\hbar\sum_{j>k}K_{jk}\,\hat{\sigma}_z(\hat{a}_j^\dagger \hat{a}_k+\hat{a}_k^\dagger \hat{a}_j),
\end{equation}
with
\begin{equation}
K_{jk}=-\frac {g_j g_k}{2}\Big(\frac{1}{\delta_j}+\frac{1}{\delta_k}\Big).
\label{eq:K_jk}
\end{equation}
Thus, after the first SW step the effective Hamiltonian is
\begin{equation}
\hat H_{\mathrm{eff}}^{(1)}
=\hat H_0+\hbar\,\hat{\sigma}_z\sum_{j}\chi_j\!\big(\hat{n}_j+\tfrac12\big)
+\hbar\sum_{j>k}K_{jk}\,\hat{\sigma}_z(\hat{a}_j^\dagger \hat{a}_k+\hat{a}_k^\dagger \hat{a}_j)
+\text{const.}
\label{eq:summary1}
\end{equation}

\paragraph{Second SW step (large detuning separations).}
Let $\delta_{jk}\coloneqq\delta_j-\delta_k$.
After the first SW step, the effective Hamiltonian reads as
Eq.~\eqref{eq:summary1},
where $\hat H_{\rm disp}$ and $\hat H_{\rm cross}$ are both of order $g^2/\delta$.
When the detuning separations satisfy $|\delta_{jk}|\gg |K_{jk}|$ for all pairs $(j,k)$,
the mixing term $\hat H_{\rm cross}$ is off-resonant in the frame set by $\hat H_0$
and can be eliminated to leading order by a second SW transformation.

We choose an anti-Hermitian generator $\hat S_2$ that solves
\begin{equation}
[\hat S_2,\hat H_0]+\hat H_{\mathrm{cross}}=0,
\end{equation}
namely
\begin{equation}
\hat S_2=\pm\sum_{j>k}\frac{K_{jk}}{\delta_{jk}}\,
\hat{\sigma}_z(\hat{a}_j^\dagger \hat{a}_k - \hat{a}_k^\dagger \hat{a}_j),
\qquad \hat S_2^\dagger=-\hat S_2 .
\label{eq:S2}
\end{equation}
Applying the unitary transformation $e^{\hat S_2}$ and expanding to second order gives
\begin{align}
\hat H_{\rm eff}^{(2)}
&\coloneqq e^{\hat S_2}\hat H_{\mathrm {eff}}^{(1)} e^{-\hat S_2} \nonumber\\
&= \hat H_0+\hat H_{\rm disp}
+[\hat S_2,\hat H_{\rm disp}]
+\frac{1}{2}[\hat S_2,\hat H_{\rm cross}]
+\cdots
\label{eq:Heff2_general}
\end{align}

\emph{(i) Commutator with the dispersive shift.}
Using $\hat H_{\rm disp}=\hbar\hat\sigma_z\sum_{j}\chi_j(\hat n_j+\tfrac12)$, one obtains the exact expression
\begin{equation}
[\hat S_2,\hat H_{\rm disp}]
=\mp \hbar\sum_{j>k}\frac{K_{jk}}{\delta_{jk}}(\chi_j-\chi_k)\,
(\hat a_j^\dagger \hat a_k+\hat a_k^\dagger \hat a_j),
\label{eq:S2_Hdisp}
\end{equation}
which is spin independent and has the form of an additional off-resonant beam-splitter
between modes $j$ and $k$.

\emph{(ii) Second-order correction from $\hat H_{\rm cross}$.}
Expanding $\tfrac12[\hat S_2,\hat H_{\rm cross}]$ yields a spin-independent differential shift.
The terms with identical pairs $(j,k)=(l,m)$ give
\begin{equation}
\Big(\frac{1}{2}[\hat S_2,\hat H_{\mathrm{cross}}]\Big)_{\mathrm{diag}}
=\pm\hbar\sum_{j>k}\frac{K_{jk}^2}{\delta_{jk}}\,(\hat{n}_j-\hat{n}_k)
+\text{const.},
\label{eq:residual_diag}
\end{equation}
while for three or more modes there are additional shared-index contributions which generate spin-independent beam-splitter interactions between other mode pairs:
\begin{equation}
\Big(\frac{1}{2}[\hat S_2,\hat H_{\mathrm{cross}}]\Big)_{\mathrm{mix}}
=\mp\frac{\hbar}{2}\sum_{k>l}\ \sum_{j\neq k,l}
K_{jk}K_{jl}\!\left(\frac{1}{\delta_{jk}}+\frac{1}{\delta_{jl}}\right)
(\hat a_k^\dagger \hat a_l+\hat a_l^\dagger \hat a_k).
\label{eq:residual_mix_compact}
\end{equation}
For two modes, the inner sum is empty and Eq.~\eqref{eq:residual_mix_compact} vanishes.

Eqs.~\eqref{eq:S2_Hdisp}--\eqref{eq:residual_mix_compact} are all spin independent.
Using $\chi\sim g^2/\delta$ and $K\sim g^2/\delta$,
their typical magnitude scales as
\begin{equation}
[\hat S_2,\hat H_{\rm disp}],\ \frac12[\hat S_2,\hat H_{\rm cross}]
\ \sim\ \hbar\,\frac{(g^2/\delta)^2}{|\delta_{jk}|}.
\end{equation}
In the regime $|\delta_{jk}|\gg |K_{jk}|$ (and typically also $|\delta_{jk}|\gg|\chi_j-\chi_k|$),
these residual terms are negligibly small compared to the leading dispersive shifts $\chi_j$,
and we neglect them in the main text.

\subsection{S1.2. Multimode nonlinear Jaynes-Cummings model (beyond Lamb-Dicke regime)}
\label{sect:derivation/nonlinear}

We now go beyond the Lamb-Dicke approximation from Eq.~\eqref{eq:Single-tone} by keeping the full displacement operators and deriving the number-dependent sideband prefactors from the exact Fourier expansion of the displacement.
For a single mode (index omitted) the displacement factor can be normal ordered as
\begin{equation}
e^{\,i\eta(\hat a e^{-i\omega t}+\hat a^\dagger e^{i\omega t})}
= e^{-\eta^2/2}\sum_{m,n\ge0}\frac{(i\eta)^{m+n}}{m!\,n!}(\hat a^\dagger)^m \hat a^n\,e^{i(m-n)\omega t}.
\end{equation}
Each exponent pair $(m,n)$ contributes to a Fourier component oscillating at frequency $(m-n)\omega$.
When we drive a single sideband and apply the rotating-wave approximation, we retain only the slowly rotating components whose Fourier indices match the chosen sideband, $n=m\pm1$ for RSB (BSB).
The terms are composed of a single annihilation (creation) operator and a polynomial of the number operator $\hat n = \hat a^\dagger \hat a$. The operator $(\hat a^\dagger)^m \hat a^m$ is diagonal in the Fock basis and equals the falling factorial of $\hat n$:
\begin{equation}
(\hat a^\dagger)^m \hat a^m=\hat n^{\underline m}\coloneqq\hat n(\hat n-1)\cdots(\hat n-m+1),
\qquad
\hat n^{\underline m}\,\ket{n}=
\begin{cases}
\dfrac{n!}{(n-m)!}\,\ket{n}, & m\le n,\\[6pt]
0, & m>n.
\end{cases}
\label{eq:S12:falling}
\end{equation}
It is convenient to introduce a nonlinear function of $\hat n$ that compactly encodes the number dependence beyond the Lamb-Dicke regime~\cite{vogelNonlinearJaynesCummingsDynamics1995, chengNonlinearQuantumRabi2018a}:
\begin{align}
f(\hat n)
&=e^{-\eta^2/2}
\sum_{m=0}^{\infty}\frac{(-\eta^2)^{m}}{m!(m+1)!}\,(\hat a^\dagger)^m \hat a^m \nonumber\\
&=e^{-\eta^2/2}
\sum_{m=0}^{\infty}\frac{(-\eta^2)^{m}}{m!(m+1)!}\frac{\hat n!}{(\hat n-m)!} \nonumber\\
&=e^{-\eta^2/2}\sum_{m=0}^{\infty}{\binom{\hat n}{m}}\frac{(-\eta^2)^{m}}{(m+1)!} \nonumber\\
&=e^{-\eta^2/2}\frac {L_{\hat n}^{(1)}(\eta^2)}{\hat n +1},
\label{eq:f(n)}
\end{align}
where $L_n^{(1)}$ is an associated Laguerre polynomial. Note that $f(\hat n )$ is a polynomial of $\hat n$.
For the single-mode case, plugging Eq.~\eqref{eq:f(n)} into Eq.~\eqref{eq:Single-tone} yields the nonlinear Jaynes-Cummings (or anti-Jaynes-Cummings) Hamiltonian (with $\phi=-\pi/2$):
\begin{equation}
\hat H_I(t)\simeq \hbar g\Big[\hat{\sigma}_\pm f(\hat n)\hat{a}\,e^{\mp i\delta t}
+\hat{\sigma}_\mp \hat{a}^\dagger f(\hat n)\,e^{\pm i\delta t}\Big],
\qquad g=\frac{\eta\Omega}{2}, \qquad  \Delta = \omega_L-\omega_0 = \mp\omega + \delta,
\end{equation}
where the upper (lower) sign corresponds to the RSB (BSB).

\medskip
\noindent
\paragraph{Multimode spectator factors.}
In a multimode setting, when mode $j$ is addressed, the spectator modes $\ell\neq j$ contribute to the diagonal factors
\begin{align}
B_\ell(\hat n_\ell)
&=e^{-\eta_\ell^2/2}
\sum_{m=0}^{\infty}\frac{(-\eta_\ell^2)^{m}}{m!\,m!}\,(\hat a_\ell^\dagger)^m \hat a_\ell^m \nonumber\\
&=e^{-\eta_\ell^2/2}\sum_{m=0}^{\infty}{\binom{\hat n_\ell}{m}}\frac{(-\eta_\ell^2)^{m}}{m!} \nonumber\\
&=e^{-\eta_\ell^2/2}\,L_{\hat n_\ell}(\eta_\ell^2),
\end{align}
and we define their product
\begin{equation}
\hat M_j\coloneqq\prod_{\ell\neq j} B_\ell(\hat n_\ell).
\label{eq:S12:spectator}
\end{equation}
With per-mode couplings and detunings
\[
g_j=\frac{\eta_j\Omega}{2},\qquad
\Delta=\omega_L-\omega_0 = \mp \omega_j + \delta_j ,
\]
the sideband-RWA interaction reads
\begin{equation}
\hat H_I(t)\simeq \hbar\sum_j g_j\Big[\hat{\sigma}_\pm \tilde{a}_j\,e^{\mp i\delta_j t}
+\hat{\sigma}_\mp \tilde{a}^\dagger_j\,e^{\pm i\delta_j t}\Big],
\end{equation}
where
\begin{equation}
\tilde{a}_j\coloneqq f_{j}(\hat n_j)\,\hat{a}_j\,\hat M_j,
\qquad
f_{j}(\hat n_j)=e^{-\eta_j^2/2}\,\frac{L_{\hat n_j}^{(1)}(\eta_j^2)}{\hat n_j+1}.
\label{eq:S12:Kdef}
\end{equation}

As in Section S1.1, we remove the explicit time dependence by the detuning-rotation frame
$\ket{\psi'(t)}=\hat W(t)\ket{\psi(t)}$ with
\begin{equation}
\hat W(t)=\exp\!\Big(\mp i t\sum_j \delta_j \hat n_j\Big),
\qquad \hat n_j=\hat a_j^\dagger \hat a_j ,
\end{equation}
where the upper (lower) sign corresponds to the RSB (BSB).
Since $\hat W$ commutes with any function of number operators, it acts on the dressed operators as
$\hat W \tilde a_j \hat W^\dagger=\tilde a_j e^{\pm i\delta_j t}$, canceling the phases in $\hat H_I(t)$.
The transformed Hamiltonian becomes time independent, $\hat H=\hat H_0+\hat V$, with
\begin{equation}
\hat H_0=\pm\hbar\sum_j \delta_j\,\hat n_j,
\qquad
\hat V=\hbar\sum_j g_j\big(\hat\sigma_\pm \tilde a_j+\hat\sigma_\mp \tilde a_j^\dagger\big).
\label{eq:S12:Hforms}
\end{equation}

\medskip
\noindent
\emph{(i) Dispersive shift (same mode).}
In the dispersive regime $|\delta_j|\gg g_j\sqrt{n_j+1}$, we repeat the SW steps of Section S1.1 after the substitution $\hat a_j\to\tilde a_j$.

We choose $\hat S_1$ again that solves $[\hat S_1,\hat H_0]+\hat V=0$:
\begin{equation}
\hat S_1=\mp\sum_{j}\frac{g_j}{\delta_j}\big(\hat{\sigma}_{\pm} \tilde{a}_j-\hat{\sigma}_{\mp} \tilde{a}_j^\dagger\big),
\qquad \hat S_1^\dagger=-\hat S_1.
\label{eq:S1_hatted}
\end{equation}
To second order, the effective Hamiltonian is
\begin{equation}
\hat H_{\mathrm{eff}}^{(1)}=\hat H_0+\frac{1}{2}[\hat S_1,\hat V].
\end{equation}
The number-shift identities
\begin{equation}
\hat a_j^\dagger f_j(\hat n_j)= f_j(\hat n_j-1)\,\hat a_j^\dagger, \qquad  f_j(\hat n_j)\,\hat a_j= \hat a_j\, f_j(\hat n_j-1),
\end{equation}
allow us to reorder diagonal functions next to the ladder operators.
Using these identities, one finds
\begin{equation}
[\hat\sigma_\pm\tilde a_j,\hat\sigma_\mp\tilde a_j^\dagger]
=\frac 1 2 \hat M_j^2\Bigg(\pm\hat \sigma _z\Big(f_j(\hat n_j)^2(\hat n_j +1)+f_j(\hat n_j-1)^2\hat n_j \Big)
+\Big(f_j(\hat n_j)^2(\hat n_j+1)-f_j(\hat n_j-1)^2 \hat n_j\Big)\Bigg).
\end{equation}
Collecting diagonal terms, we obtain the spin-dependent and spin-independent parts of the dispersive shift as Eq.~\eqref{eq:H_disp}:
\begin{equation}
\hat H_\mathrm{disp,\,spin\text{-}dep}
=\hbar\sum_{j}\frac {\chi_j} 2  \hat \sigma_z\,\hat M_j^2
\Big(f_j(\hat n_j)^2(\hat n_j+1)+f_j(\hat n_j-1)^2\hat n_j\Big),
\label{eq:disp, spin-dep}
\end{equation}
\begin{equation}
\hat H_\mathrm{disp,\,spin\text{-}indep}
=\pm\hbar\sum_{j}\frac {\chi_j} 2 \,\hat M_j^2
\left(
f_j(\hat n_j)^2(\hat n_j+1)-f_j(\hat n_j-1)^2 \hat n_j
\right),
\label{eq:disp,spin-indep}
\end{equation}
where $\chi_j=-g_j^2/\delta_j$.

The term $\hat H_\mathrm{disp,\,spin\text{-}indep}$ is diagonal in the Fock basis and generates a purely motional, phonon-number-dependent phase without changing the Fock state populations.
Therefore it can be neglected for population-based analyses and single-shot Fock state measurement protocols that depend only on $\{p_{\boldsymbol n}\}$.
It also does not directly affect the outcome probabilities of parity measurements, which are diagonal in the number basis.
However, in parity-filtering protocols where the postselected motional state is used as a reusable resource (e.g., for preparing cat states or entangled coherent states), $\hat H_\mathrm{disp,\,spin\text{-}indep}$ gives additional number-dependent phases during the filtering sequence, which can distort the relative phases between Fock components and reduce the fidelity to the ideal target state unless compensated.

\medskip
\noindent
\emph{(ii) Spin-dependent beam splitter (pairwise mixing between different modes).}
The first SW step also generates pairwise mode mixing.
It is most transparent to express the result in terms of the dressed ladder operators
$\tilde a_j=f_j(\hat n_j)\hat a_j\hat M_j$ introduced in Eq.~\eqref{eq:S12:Kdef}.
We obtain
\begin{equation}
\hat H_{\mathrm{cross}}
=\hbar\sum_{j>k}K_{jk}\,\hat\sigma_z\,
\big(\tilde a_j^\dagger \tilde a_k+\tilde a_k^\dagger \tilde a_j\big),
\label{eq:H_cross_nonlinear}
\end{equation}
with $K_{jk}=-\frac{g_j g_k}{2}\left(\frac{1}{\delta_j}+\frac{1}{\delta_k}\right)$ as in Eq.~\eqref{eq:K_jk}.
In the Lamb-Dicke limit, $f_j(\hat n_j)\to 1$ and $\hat M_j\to 1$, so Eq.~\eqref{eq:H_cross_nonlinear} reduces to the linear spin-dependent beam splitter.

When the detuning separations $|\delta_{jk}|\coloneqq|\delta_j-\delta_k|$ are large compared to $|K_{jk}|$,
$\hat H_{\mathrm{cross}}$ is off-resonant in the frame set by $\hat H_0$ and can be eliminated by a second SW step
(see Section S1.1).
The remaining terms are spin independent and appear at the same order from
$[\hat S_2,\hat H_{\rm disp}]$ and $\tfrac12[\hat S_2,\hat H_{\rm cross}]$,
with a typical magnitude $\sim \hbar (g^2/\delta)^2/|\delta_{jk}|$.

In the following, we neglect these residual spin-independent corrections and focus on the spin-dependent diagonal dispersive term
$\hat H_{\mathrm{disp,\,spin\text{-}dep}}$ as the dominant source of phonon-number-dependent phases in a Ramsey experiment.
For a general initial motional state $\hat\rho_{\rm mot}$, the Ramsey signal depends only on its diagonal Fock populations
\[
p_{\boldsymbol n}\coloneqq\bra{\boldsymbol n}\hat\rho_{\rm mot}\ket{\boldsymbol n},
\qquad \sum_{\boldsymbol n}p_{\boldsymbol n}=1,
\]
and the probability of finding the spin in $\ket{\uparrow}$ is
\begin{equation}
P_{\uparrow}(t)
= \tfrac{1}{2}-\tfrac{1}{2}\sum_{\boldsymbol n} p_{\boldsymbol n}\,
\cos\!\big[\Phi_{\boldsymbol n}(t)-\phi_{\mathrm{off}}(t)\big].
\label{eq:Pr-up-cos-nonlinear_rho}
\end{equation}
where $\phi_{\mathrm {off}}$ is the residual shift offset compensation as described in Section S4.1. The relative number-dependent nonlinear phase between the spin states $\ket{\uparrow}$ and $\ket{\downarrow}$ is
\begin{align}
\Phi_{\boldsymbol n}(t)
&= t\sum_{j}\chi_j\,M_j(\boldsymbol n)^{2}\,
\Big(\,f_j(n_j)^2(n_j+1)+f_j(n_j\!-\!1)^2\,n_j \Big) \nonumber\\
&= t\sum_{j}\chi_j\,M_j(\boldsymbol n)^{2}\,e^{-\eta_j^2}\,
\left(
 \frac{\big[L_{n_j}^{(1)}(\eta_j^2)\big]^2}{n_j+1}
+\frac{\big[L_{n_j-1}^{(1)}(\eta_j^2)\big]^2}{n_j}
\right).
\label{eq:Phi-nonlinear}
\end{align}
Here
\begin{equation}
M_j(\boldsymbol n)=\prod_{\ell\neq j} e^{-\eta_\ell^2/2}\,L_{n_\ell}(\eta_\ell^2),
\qquad
f_j(n_j)=e^{-\eta_j^2/2}\,\frac{L_{n_j}^{(1)}(\eta_j^2)}{n_j+1}.
\end{equation}
The second term in parentheses in Eq.~\eqref{eq:Phi-nonlinear} is zero for $n_j=0$ since $f_j(n_j-1)$ is a polynomial of $n_j$. Thus the nonlinearity is fully captured by the Laguerre polynomials $L_n$ and $L_n^{(1)}$.

\medskip
\noindent\emph{Consistency (Lamb-Dicke limit).}
Expanding $f_j(n)=1-\tfrac{1}{2}(n+1)\eta_j^2+\cdots$ and
$M_j(\boldsymbol n)^2=1-\sum_{\ell\neq j}(2n_\ell+1)\eta_\ell^2+\cdots$,
Eq.~\eqref{eq:Phi-nonlinear} reduces to
\begin{equation} \label{eq:Phi-leading-order}
\Phi_{\boldsymbol n}(t)=t\sum_j \chi_j(2n_j+1)+\mathcal O(\eta^2).
\end{equation}
Discarding the global, number-independent phase recovers the linear result
\begin{equation} \label{eq:P_uparrow}
P_{\uparrow}(t)
=\tfrac{1}{2}-\tfrac{1}{2}\sum_{\boldsymbol n}p_{\boldsymbol n}
\cos\!\big(\boldsymbol\theta(t)\!\cdot\!\boldsymbol n\big),
\qquad \theta_j(t)=2\chi_j t,
\end{equation}
which matches Eq.~(6) in the main text.

\section{S2. data analysis}

\subsection{S2.1. Analytic nonlinear function $P_\uparrow(t)$ for curve fitting}
To obtain more accurate Fock-state population, we consider (\romannumeral 1) the nonlinearity beyond the Lamb-Dicke regime, (\romannumeral 2) the step-averaged effect (yielding the effective dispersive shifts), (\romannumeral 3) the phonon-number-dependent exponential decay of the contrast, and (\romannumeral 4) the residual shift offset when fitting the experimental data. Since Eq.~\eqref{eq:Phi-nonlinear} already incorporates (\romannumeral 1), we use them to calculate the total phase accumulation under the selective decoupling scheme.

\paragraph{Calculation of the effective dispersive shift.}
The two-step Ramsey sequence with a spin echo for the selective decoupling cancels the contribution of the carrier AC-Stark shift, while preserving that of the phonon-number-dependent shift. Denoting the phonon-number-dependent nonlinear phases within step $k$ as $\Phi_{\boldsymbol n}^{(k)}$, the total phase after two steps is
\begin{equation} \label{eq:Phi-total}
    \Phi_{\mathrm{tot}, \boldsymbol{n}}=-\Phi_{\boldsymbol n}^{(1)} + \Phi_{\boldsymbol n}^{(2)},
\end{equation}
where the spin echo negates the sign of $\Phi_{\boldsymbol n}^{(1)}$. Note that, Eq.~\eqref{eq:Phi-nonlinear} is a linear combination of $\chi_j$'s and the coefficients depends only on the phonon numbers $\boldsymbol{n}$. For brevity, we define $S_j(\boldsymbol{n})$ to be the coefficients, so that
\begin{equation}
    \Phi_{\boldsymbol{n}}(t) = t \sum_j \chi_j S_j(\boldsymbol{n}).
\end{equation}
Now we can rewrite Eq.~\eqref{eq:Phi-total} with total-time dependence as
\begin{equation}
\begin{split}
    \Phi_{\mathrm{tot}, \boldsymbol{n}} (t) & = \sum_j (-t^{(1)}\chi_j^{(1)} + t^{(2)}\chi_j^{(2)}) S_j(\boldsymbol{n}) \\
    & = t \sum_j \chi_{\mathrm{eff}, j} S_j(\boldsymbol{n}),
\end{split}
\end{equation}
where $t=t^{(1)} + t^{(2)}$ is the total interaction time, so $t^{(k)}=t \Delta^{\!(k)} / (\Delta^{\!(1)}+\Delta^{\!(2)})$. The effective dispersive shift $\chi_{\mathrm{eff}, j} = \sum_k (-1)^k t^{(k)} \chi_j^{(k)} / \sum_{k'} t^{(k')}$ is defined in the main text as well.

\paragraph{Phonon-number-dependent exponential decay of the contrast.}
As denoted in Eq.~(11) in the main text, the spin oscillation has an exponential decay in the contrast. We model the decay rate as linear in the phonon numbers:
\begin{equation}
    \gamma_{\boldsymbol{n}} = \sum_j \gamma_j (2n_j +1),
\end{equation}
because a linear decay model is previously used~\cite{wollackQuantumStatePreparation2022}, and we assume that the linearity stems from the phonon-number-dependent phase, as shown in Eq.~\eqref{eq:Phi-leading-order}.

\paragraph{Residual shift offset.}
Using the phase of the second $\pi/2$ pulse in the Ramsey sequence, we nullify the phase accumulation of the motional vacuum $\ket{0}$. However, this may fail to completely suppress the residual shift, probably due to the fluctuation of the beam intensity and the repetition rate of the pulse laser, which result in a drifting carrier AC-Stark shift. If there is a remaining shift offset $\chi_\mathrm{res} \hat \sigma_z$, then the total phase acquires an additional constant phase accumulation
\begin{equation}
    \Phi_{\mathrm{res}}(t)=2\chi_{\mathrm{res}}t.
\end{equation}

\paragraph{Fitting function.}
Considering all the effects discussed above, we design the function for fitting the spin dynamics as follows:
\begin{equation} \label{eq:g_uparrow}
    g_\uparrow(t;\gamma_{j}, \chi_{\mathrm{eff}, j}, \chi_\mathrm{res}) = \sum_{\boldsymbol{n}} \frac{p_{\boldsymbol{n}}}{2}\left(1 - e^{-\gamma_{\boldsymbol{n}}t}\cos{(\Phi_{\mathrm{tot}, \boldsymbol{n}}(t) + \Phi_{\mathrm{res}}(t) - \phi_{\mathrm{off}}(t))}\right),
\end{equation}
where $p_{\boldsymbol{n}}=\bra{\boldsymbol{n}} \hat \rho_{\mathrm{mot}} \ket{\boldsymbol{n}}$ is the population of the Fock state $\ket{\boldsymbol{n}}$, given the density matrix of the motional state $\hat \rho_{\mathrm{mot}}$.

We reduce the number of free parameters by fixing the ratio $r \coloneqq \chi_{\mathrm{eff}, 1} / \chi_{\mathrm{eff}, 2}=\gamma_1 / \gamma_2$ to the target value, because we independently calibrate this ratio using Fock states. Note that we let the ratio of the decay rates be the same as the ratio of the effective dispersive shifts, as we already assumed that the decay rates originate from the phonon-number-dependent phase. Then we fit three free parameters: $\gamma_1$, $\chi_{\mathrm{eff}, 1}$, $\chi_{\mathrm{res}}$, along with the Fock-state population $p_{\boldsymbol{n}}$. We limit the Hilbert space dimension to $n_{\mathrm{max}} + 1$ for each mode, yielding $(n_{\mathrm{max}} + 1)^2 + 3$ free parameters in total in the two-mode setting. In the single-mode setting, the total number of free parameters is $n_\mathrm{max}+4$, and we set $\gamma_2$ and $\chi_{\mathrm{eff},2}$ to zeros. We set the lower bounds to zero for all the parameters except $\chi_{\mathrm{res}}$. We do not give any penalty or constraints to the summation of the populations $\sum_{\boldsymbol{n}} p_{\boldsymbol{n}}$, which should be nearly unity in the ideal case---the summation only includes the Fock numbers that satisfy $\max_j n_j \leq n_{\mathrm{max}}$. Instead, we check if the summation is close to 1. Experimentally, the total population may deviate from unity due to the suboptimal state detection, which rescales the value of $g_\uparrow(t)$. However, when estimating the parity, we normalize the population. The fitted parameter values for Figs.~5 and 6 in the main text are compiled in Table~\ref{tab:fitted-parameters-main-text}. $n_\mathrm{max}$ is set to 6 and 4 for the single-mode and two-mode cases, respectively.

\begin{table}[h]
    \centering
    \begin{ruledtabular}
    \begin{tabular}{ccdddddd}
    \textrm{Figure}&
    \textrm{Ratio $r$}&
    \multicolumn{1}{c}{\textrm{$\gamma_1 \ \mathrm{(s^{-1})}$}}&
    \multicolumn{1}{c}{\textrm{$\chi_{\mathrm{eff,1}} / (2\pi) \ \mathrm{(Hz)}$}}&
    \multicolumn{1}{c}{\textrm{$\chi_{\mathrm{res}} / (2\pi) \ \mathrm{(Hz)}$}}&
    \multicolumn{1}{c}{\textrm{$\sum_{\boldsymbol{n}} p_{\boldsymbol{n}}$}}&
    \multicolumn{1}{c}{\textrm{Ideal $\sum_{\boldsymbol{n}} p_{\boldsymbol{n}}$}}&
    \multicolumn{1}{c}{\textrm{Parity}}\\
    \colrule
    \colrule
    5(a) &  & 15(4) & -174(2) & 2(5) & 0.95(4) & 0.9964 & 0.90(5)\\
    \colrule
    5(b) &  & 9(3) & -170(1) & 3(2) & 0.94(3) & 0.9867 & -0.73(4)\\
    \colrule
     & 0.5 & 11(2) & -106(1) & -7(3)  \\
    6(a) & 1 & 10(4) & -203(2) & -9(5) & 0.98(4) & 0.9935 & 0.73(6)\\
     & 2 & 19(5) & -254(3) & -4(4)  \\
    \colrule
     & 0.5 & 12(3) & -106(2) & -7(3)  \\
    6(b) & 1 & 15(5) & -209(2) & -6(3) & 0.97(4) & 0.9918 & -0.65(7)\\
     & 2 & 9(5) & -251(2) & 0(2)  \\
    \end{tabular}
    \end{ruledtabular}
    \caption{Fitted parameter values for the data shown in the Figs. 2 and 3 in the main text. The ideal population coverage $\sum_{\boldsymbol{n}} p_{\boldsymbol{n}}$ is calculated for the ideal cat states and ECSs with target $\alpha=1.5$ and $1.0$, respectively.}
    \label{tab:fitted-parameters-main-text}
\end{table}

\begin{table}[h]
    \centering
    \begin{ruledtabular}
    \begin{tabular}{ccdddddd}
    \textrm{Figure}&
    \textrm{Ratio $r$}&
    \multicolumn{1}{c}{\textrm{$\gamma_1 \ \mathrm{(s^{-1})}$}}&
    \multicolumn{1}{c}{\textrm{$\chi_{\mathrm{eff,1}} / (2\pi) \ \mathrm{(Hz)}$}}&
    \multicolumn{1}{c}{\textrm{$\chi_{\mathrm{res}} / (2\pi) \ \mathrm{(Hz)}$}}&
    \multicolumn{1}{c}{\textrm{$\sum_{\boldsymbol{n}} p_{\boldsymbol{n}}$}}&
    \multicolumn{1}{c}{\textrm{Ideal $\sum_{\boldsymbol{n}} p_{\boldsymbol{n}}$}}&
    \multicolumn{1}{c}{\textrm{Parity}}\\
    \colrule
    \colrule
    \ref{fig:sdf-cat}(a) &  & 11(3) & -178(1) & -12(2) & 0.97(3) & 0.9964 & 0.92(4)\\
    \colrule
    \ref{fig:sdf-cat}(b) &  & 14(3) & -188(1) & -1(2) & 0.98(4) & 0.9867 & -0.93(5)\\
    \colrule
     & 0.5 & 1(2) & -108(1) & -16(2)  \\
    \ref{fig:sdf-ecs}(a) & 1 & 9(4) & -236(4) & -8(8) & 0.99(3) & 0.9935 & 0.84(4)\\
     & 2 & 14(6) & -282(3) & -8(5)  \\
    \colrule
     & 0.5 & 5(3) & -107(1) & -4(2)  \\
    \ref{fig:sdf-ecs}(b) & 1 & 10(4) & -243(2) & -3(3) & 0.99(3) & 0.9918 & -0.71(5)\\
     & 2 & 23(6) & -285(3) & -8(3)  \\
    \end{tabular}
    \end{ruledtabular}
    \caption{Fitted parameter values for the data shown in the Figs. \ref{fig:sdf-cat} and \ref{fig:sdf-ecs}. The ideal population coverage $\sum_{\boldsymbol{n}} p_{\boldsymbol{n}}$ is calculated for the ideal cat states and ECSs with target $\alpha=1.5$ and $1.0$, respectively.}
    \label{tab:fitted-parameters-supplemental}
\end{table}

\subsection{S2.2. Alternative generation method of cat states and ECSs}
\begin{figure}
    \centering
    \includegraphics[width=0.5\linewidth]{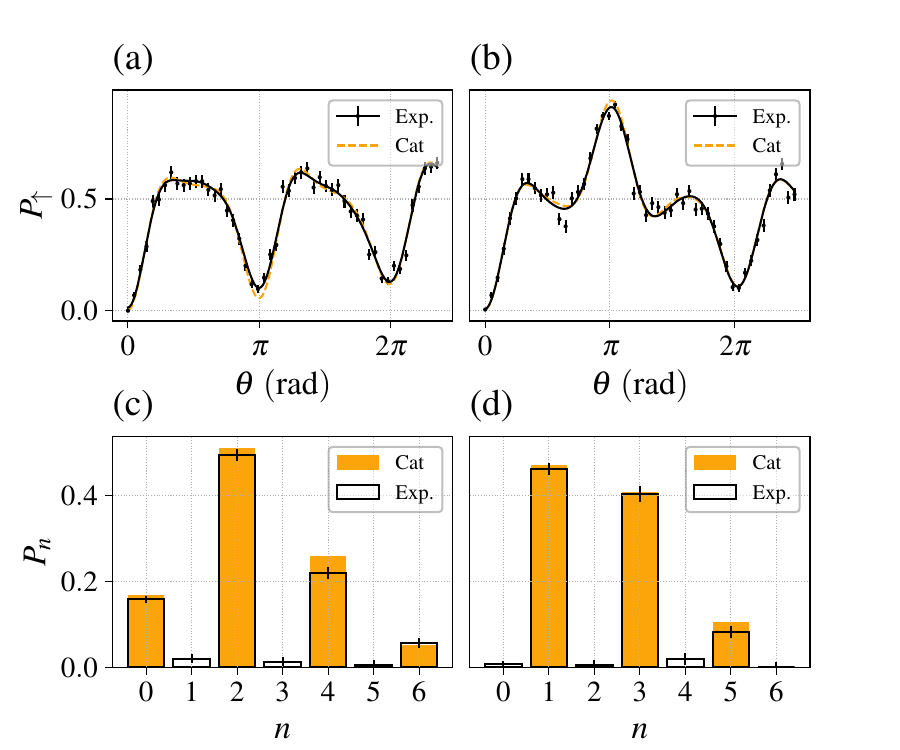}
    \caption{Fock-state population fitting for the single-mode cat states $\ket{\alpha}\pm\ket{-\alpha}$. (a), (b) Time evolutions of spin population under the Ramsey sequence for the even- and odd-cat states, respectively, where $\theta=\chi_{\mathrm{eff},1}\sum_k t^{(k)}$. The markers denote experimental data and the solid curves are fits, and the orange dashed curves show the expected curves for ideal cat states. (c), (d) Corresponding Fock-state populations extracted from the fits (black open rectangles). The orange bars show the Fock-state distributions of the best-fit ideal even- and odd-cat states, respectively. The fitted $\alpha$ for the even- and odd-cat states are 1.57(1) and 1.51(1), respectively.}
    \label{fig:sdf-cat}
\end{figure}

\begin{figure}
    \centering
    \includegraphics[width=0.5\linewidth]{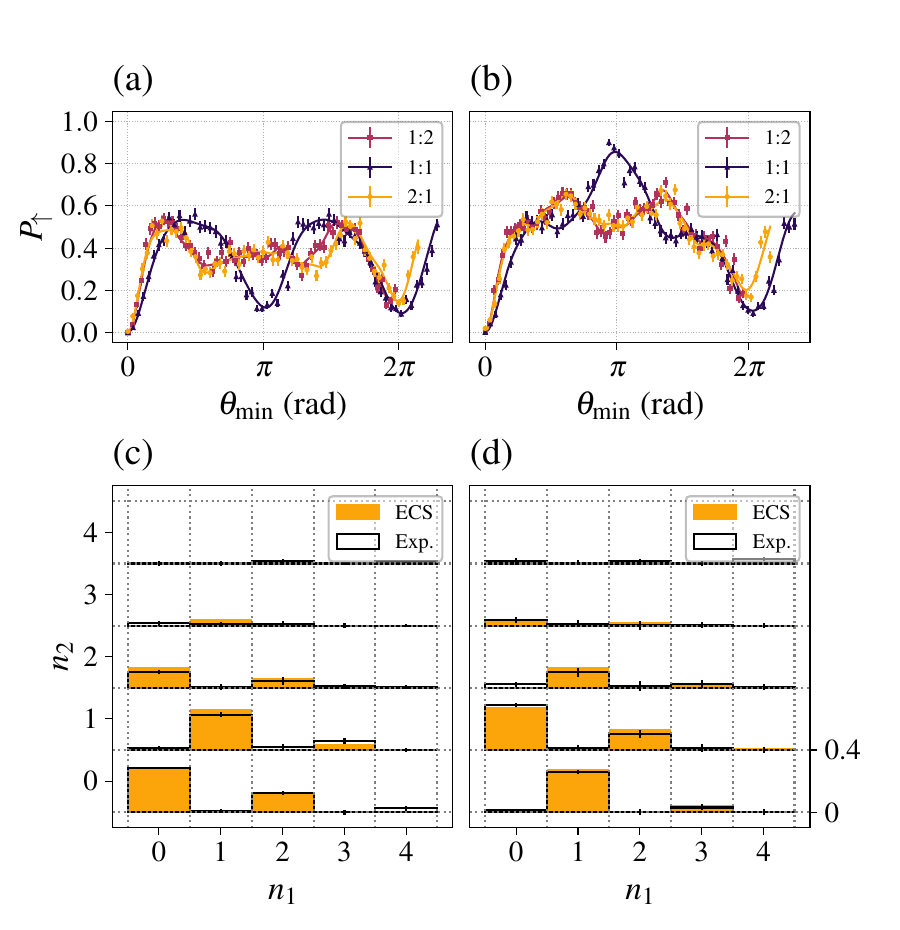}
    \caption{Fock-state population fitting for the two-mode ECSs $\ket{\alpha, \alpha}\pm\ket{-\alpha, -\alpha}$. (a), (b) Time evolutions of spin population under the Ramsey sequence for the even and odd ECSs, respectively, where $\theta_\mathrm{min}=\min(\chi_{\mathrm{eff},1}, \chi_{\mathrm{eff},2})\sum_k t^{(k)}$. The markers denote experimental data and the solid curves are fits. The legend labels indicate the ratio $\chi_{\mathrm{eff},1}:\chi_{\mathrm{eff},2}$. (c), (d) Corresponding Fock-state populations extracted from the fits (black open rectangles). The orange bars show the Fock-state distributions of the best-fit ideal even and odd ECSs, respectively. A single grid height corresponds to the phonon population of 0.4. The fitted $\alpha$ for the even and odd ECSs are 0.98(1) and 0.99(2), respectively.}
    \label{fig:sdf-ecs}
\end{figure}

In the main text, we show the measured Fock-state population of single-mode cat states and two-mode entangled coherent states (ECSs), which are generated by applying the parity-based filtering on coherent states. On the other hand, we can also directly prepare those states using the spin-dependent force (SDF) with the superposition of its spin eigenstates, and then postselecting on a spin state~\cite{jeonMultimodeBosonicState2025}. This alternative method is simpler to implement experimentally, hence expected to present a better state generation fidelity. Therefore, we measure the Fock-state population of the directly-prepared cat states and ECSs to demonstrate the dispersive shift used solely for the Fock-state population measurement, without the parity-based filtering steps. In Figs.~\ref{fig:sdf-cat} and \ref{fig:sdf-ecs} we show the results for the single-mode cat states and the two-mode ECSs, respectively. The fitted parameter values are compiled in Table~\ref{tab:fitted-parameters-supplemental}. In fact the estimated parities show better contrast, compared to Table~\ref{tab:fitted-parameters-main-text}.

\subsection{S2.4. Experimental parameters}
In this section, we enumerate the actual experimental parameters used for each setting. There are four experimental settings in this work: single-mode, multimode $1:2$, multimode $1:1$, and multimode $2:1$, where the multimode settings denote their effective dispersive shift ratio $\chi_{\mathrm{eff}, 1} : \chi_{\mathrm{eff}, 2}$. For each setting, the detunings from the blue sidebands ($\delta_j^{(k)}=\Delta^{\!(k)}-\omega_j$) are compiled in Table~\ref{tab:detuning-settings}. In practice, however, the blue sideband transitions are not resonant at detunings of $\Delta=\omega_j$ due to the carrier AC-Stark shift. Therefore, we measure the mode frequencies $\omega_j$ using spin-dependent force (SDF), for which the carrier AC-Stark shift is canceled by the bichromatic beams, and use it when calculating the detunings. The beam intensity is calibrated at the carrier Rabi frequency of $100 \ \mathrm{kHz}$.

\begin{table}[h]
    \centering
    \begin{ruledtabular}
    \begin{tabular}{ccdcd}
    \textrm{Setting}&
    \textrm{Mode $j^{(1)}$}&
    \multicolumn{1}{c}{\textrm{Detuning $\delta_{j^{(1)}}^{(1)} / (2\pi) \ \mathrm{(kHz)}$}}&
    \textrm{Mode $j^{(2)}$}&
    \multicolumn{1}{c}{\textrm{Detuning $\delta_{j^{(2)}}^{(2)} / (2\pi) \ \mathrm{(kHz)}$}}\\
    \colrule
    Single-mode & 1 & 110 & 1 & -110 \\
    Multimode 1:2 & 2 & 49 & 1 & -142.5 \\
    Multimode 1:1 & 2 & 50 & 1 & -62.5 \\
    Multimode 2:1 & 2 & 93 & 1 & -50
    \end{tabular}
    \end{ruledtabular}
    \caption{Sideband detunings used in step 1 and 2 for each setting.}
    \label{tab:detuning-settings}
\end{table}

\bibliography{dispersive_shift_ref}